\documentclass[12pt]{article}
\usepackage{graphicx}
\usepackage{lineno}
\usepackage{amsfonts}
\usepackage{amsmath}
\usepackage[linesnumbered,ruled,vlined]{algorithm2e}
\usepackage{algcompatible}
\usepackage{caption}

\modulolinenumbers[5]

\newtheorem{thm}{Theorem}[section]
\newtheorem{cor}[thm]{Corollary}
 \newtheorem{lem}[thm]{Lemma}
 \newtheorem{prop}[thm]{Proposition}

\begin{document}

\title{Comparative study of space filling curves for cache oblivious TU Decomposition
}


\author{Fatima K. Abu Salem         \and
        Mira Al Arab 
}


\author{Fatima K. Abu Salem \\
              Computer Science Department \\
              American University of Beirut\\
              Tel.: +961-1-350000 ext. 4224\\
              Fax: +961-1-744461 \\
              fatima.abusalem@aub.edu.lb           
           \and
           Mira Al Arab \\
              Computer Science Department \\
              American University of Beirut\\
              maa75@aub.edu.lb
}

\date{Received: date / Accepted: date}

\maketitle

\begin{abstract}
We examine several matrix layouts based on space-filling curves that allow for a cache-oblivious adaptation of parallel TU decomposition for rectangular matrices over finite fields. The TU algorithm of \cite{Dumas} requires index conversion routines for which the cost to encode and decode the chosen curve is significant. Using a detailed analysis of the number of bit operations required for the encoding and decoding procedures, and filtering the cost of lookup tables that represent the recursive decomposition of the Hilbert curve, we show that the Morton-hybrid order incurs the least cost for index conversion routines that are required throughout the matrix decomposition as compared to the Hilbert, Peano, or Morton orders. The motivation lies in that cache efficient parallel adaptations for which the natural sequential evaluation order demonstrates lower cache miss rate result in overall faster performance on parallel machines with private or shared caches, on GPU's, or even cloud computing platforms. We report on preliminary experiments that demonstrate how the TURBO algorithm in Morton-hybrid layout attains orders of magnitude improvement in performance as the input matrices increase in size. For example, when $N = 2^{13}$, the row major TURBO algorithm concludes within about 38.6 hours, whilst the Morton-hybrid algorithm with truncation size equal to $64$ concludes within 10.6 hours.
\end{abstract}

\section{Introduction}

Exact triangulisation of matrices is crucial for a large range of problems in Computer Algebra and Algorithmic Number Theory, where a basis of the solution set of the associated linear system is required. A known algorithm in the field which resulted in several prominent adaptations is the TURBO algorithm of Dumas et al. \cite{Dumas} for exact LU decomposition. This algorithm recurses on rectangular and potentially singular matrices, which makes it possible to take advantage of cache effects. It improves on other expensive methods for handling singular matrices, which otherwise have to dynamically adjust the submatrices so that they become invertible. Particularly, TURBO significantly reduces the volume of communication on distributed architectures, and retains optimal work and linear span. TURBO can also compute the rank in an exact manner. As benchmarked against some of the most efficient exact elimination algorithms in the literature, TURBO incurs low synchronisation costs and reduces the communication cost featured by \cite{Ibarra80,Ibarra82} by a factor of one third when used with only one level of recursion on 4 processors. In TURBO, local TU factorisations are performed until the sub-matrices reach a given threshold, and so one can take advantage of cache effects. A cache friendly adaptation of the serial version of TURBO bears impact on all possible forms of parallel or distributed deployment of the algorithm. For one, nested parallel algorithms with low depth and for which the natural sequential execution has low cache complexity will also attain good cache complexity on parallel machines with private or shared caches \cite{BlellAl.10}. Locality of reference on distributed systems is also being advocated by the Databricks group initiated by founders of Apache Spark. In their own terms, when profiling Spark user applications on distributed clusters, a large fraction of the CPU time was spent waiting for data to be fetched from main memory. Locality of reference is also of concern on GPUs. Although one does not have full control over optimising locality of reference on such machines, and despite that GPUs rely on thread-level parallelism to hide long latencies associated with memory access, the memory hierarchy remains critical for many applications. Finally, applications that are cache aware are also deemed to be more energy aware, as remarked by the Green computing community.


This preamble motivates our work on trying to improve the cache performance of the serial version of TURBO. It is well established that traditional row-major or column-major layouts of matrices in compilers lead to extremely poor temporal and spatial locality of matrix algorithms. Instead, several matrix layouts based on space filling-curves have yielded cache-oblivious adaptations of matrix algorithms such as matrix-matrix multiplication \cite{Peano:Mult,Chatterjee:Layouts2} and matrix factorisation \cite{Peano:MatrixOps,Wise:QR,Wise:Cholesky}. Those alternative layouts are recursive in nature and produce highly cache-efficient, cache-oblivious adaptations that scale with the size of the underlying matrices. The cache-oblivious model does not require knowledge of, and hence tuning the algorithm according to, the cache parameters. Cache-oblivious programs allow for resource usage not to be programmed explicitly, and for algorithms to be portable across varying architectures, as well as all levels of the memory hierarchy within one specific architecture.

Our contributions can be summarised as follows:
\begin{enumerate}
\item{We investigate prospects for a cache oblivious adaptation of the TURBO algorithm by mapping four different matrix layouts against each other: the Hilbert order \cite{Hilbert:EncDecNew,Hilbert:EncDec}, the Peano order \cite{Peano:MatrixOps,Peano:Mult}, the Morton order \cite{Morton:EncDec,DilatedInts}, and the Morton-hybrid order \cite{Wise:MaskedIntegers}. Whilst matrices on which we want to perform matrix-matrix multiplication or LU decomposition without pivoting can be serialized no matter what layout is used, the recursive TU decomposition considered in this work consistently requires permutation steps that require one to traverse the matrix in a row-wise or column-wise manner, thus eliciting index conversion from the Cartesian scheme to the recursive scheme and vice versa. In addition to the specific contributions summarised below, this survey component of our work }
\item{Our analysis of the four schemes addresses the cost of bit operations and accessing table lookups when applicable. Our findings show the following:
\begin{enumerate}
\item{The overhead for using the Peano layout will be compelling as index conversion invokes operations modulo 3.}
\item{Whilst the Hilbert layout has been promising for improving memory performance of matrix algorithms in general, and despite that the operations for encoding and decoding in this layout can be performed using bit shifts and bit masks, we will still require $m$ iterations for a $2^{m} \times 2^{m}$ matrix for each single invocation of encoding or decoding.}
\item{In contrast, we find that the conversions for the Morton and the Morton-hybrid layouts incur a constant number of operations assuming the matrix is of dimensions at most $2^{\alpha} \times 2^{\alpha}$, where $\alpha$ is the machine word-size. For the typical value $\alpha = 64$, such matrix sizes are sufficiently large for many applications. }
\item{Furthermore, despite that the Morton order can be encoded and decoded faster than the Morton-hybrid order, the factor of improvement is constant: ten less operations. In return, the Morton-hybrid layout allows the recursion to stop when the blocks being divided are of some prescribed size equal to $T \times T$ , thus decreasing the recursion overhead. These $T \times T$ blocks are stored in a row-major order, which allows for benefiting from compiler optimizations that have already been designed for this layout. The row-major ordering of the block at the base case also makes accessing the entries within the blocks at the base case of the inversion, multiplication, and decomposition steps of the algorithm faster and easier because no index conversion is required.}
\end{enumerate}
}
\item{Unless otherwise stated and explicitly cited, the various encoding and decoding algorithms we present under these various layouts and the propositions/proofs associated with them, are novel.}
\item{The present manuscript is an indispensable precursor for our work in \cite{AA16b}, where we introduce the concepts of {\it alignment} of sub-matrices with respect to the cache lines and their {\it containment} within proper blocks under the Morton-hybrid layout, and describe the problems associated with the recursive subdivisions of TURBO under this scheme. Although the full details of the resulting algorithm are beyond the scope of this paper, we report on experiments that demonstrate how the TURBO algorithm in Morton-hybrid layout attains orders of magnitude improvement in performance as the input matrices increase in size. For example, when $N = 2^{13}$, the row major TURBO algorithm concludes within about 38.6 hours, whilst the Morton-hybrid algorithm with truncation size equal to $64$ concludes within 10.6 hours.}
\end{enumerate}

\section{The TU Algorithm: A Summary}
\label{TU-Summary}

Consider a $2m \times 2n$ matrix $A$ with rank $r$ over a field $\mathbb{F}$, where $A$ may be singular. The TURBO algorithm triangulates the matrix $A$ in a succession of recursive steps, relaxing the condition for generating a strictly lower triangular matrix. All the recursive steps are independent and thus TURBO is inherently parallel. It further outputs two matrices $T$ and $U$, such that $A=T \cdot U$, where $U$ is a $2m \times 2n$ upper triangular matrix, and $T$ is $2m \times 2m$, with some ``$T$'' patterns. In all of the following, let $A^{(i,j)}_{_{(a,b)}}$ denote the sub-matrix of $A$ of dimensions $a \times b$ and starting at the entry of Cartesian index $(i,j)$. Whenever the superscript $(i,j)$ is omitted, a default value $(0,0)$ is assumed. First, begin by decomposing the matrix $A$ of size $2m \times 2n$ as follows:
\begin{equation}
\label{eqn:lu} A_{_{(2m,2n)}} = \left(
\begin{array}{cc}
NW_{_{(m,n)}} & NE_{_{(m,n)}}\\
SW_{_{(m,n)}} & SE_{_{(m,n)}}
\end{array}
\right)
\end{equation}
The TURBO alogrithm now performs the following: 
\begin{enumerate}
\item{Recursive TU decomposition in each of SE, then SW, NE, and finally, NW}
\item{Virtual row and column permutations needed to re-order the blocks to yield a final, upper triangular matrix}
\end{enumerate}
For brevity, we only elaborate on the first step in the TURBO algorithm. It gives a flavour of the various matrix opertions we will be addressing throughout the paper. The rest of the steps can be found in \cite{Dumas}.
\subsection{Step 1: Recursive TU in $NW$}
\label{sec:alg_step1} 
\begin{figure}[h]
\centering
    {\includegraphics[width=4.5in]{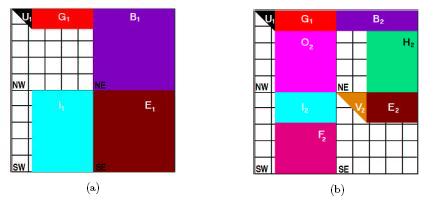}}
    \caption{Matrix after Step 1 (a) and Step 2 (b)}
\label{fig:step1_2}
\end{figure}

This step performs a recursive
call to the TU decomposition algorithm in the $NW$ quadrant of $A$ to get $U_{1}$ upper triangular, $G_1$, and $L_1$, lower triangular, such that 
\begin{equation*}
L_{1} \cdot NW = \left(\begin{array}{cc}U_{1} & G_{1}\\0&0
\end{array} \right).
\end{equation*}
Here, $U_1$ is $r \times r$ for some $r \ge 0$, $G_{1}$, and $L_{1}$ is $m \times m$. $L_{1}$ is then used to update $NE$ by:
\begin{equation*}B_{1_{(m,n)}} = L_{1} \cdot NE.\end{equation*}
One now aims to zero out the sub-matrix of SW that lies under $U_1$. This is done by calculating 
\begin{equation*}N_{1_{(m,r)}} = - SW_{_{(m,r)}} \cdot U_{1}^{-1}\end{equation*}
and then setting 
\begin{equation*} \left(
\begin{array}{cc}
0_{_{(m,r)}} & I_{1_{(m,n-r)}}
\end{array} \right) = SW + N_{1_{(m,r)}} \cdot \left(\begin{array}{cc}U_{1_{(r,r)}} & G_{1_{(r,n-r)}}.
\end{array} \right)\end{equation*}
The submatrix $SE$ has to be updated accordingly:
\begin{equation*} E_{1} = SE + N_{1} \cdot B_{1}.\end{equation*}
At the end of this step, matrix $A$ has been updated as follows:
\begin{equation*}
A_{1} = \left(
\begin{array}{ccc}
U_{1_{(r,r)}}       & G_{1_{(r,n-r)}}    & B_{1_{(r,n)}}\\
0_{_{(m-r,r)}}      & 0_{_{(m-r,n-r)}}   & B_{1_{(m-r,n)}}\\
0_{_{(m,r)}}        & I_{1_{(m,n-r)}}    & E_{1_{(m,n)}}\\
\end{array}
\right).
\end{equation*}
The resulting matrix can be seen in Fig. \ref{fig:step1_2}(a) taken
from \cite{Dumas}.

\section{Comparison of Index Conversion Overhead}
\label{:index_conversion}

The amount of computation required for index conversion from each of the Hilbert, Peano, Morton, and Morton-hybrid layouts to the Cartesian order is used to rate each of these 
layouts as they apply to TURBO. The row and column permutations required at every level
of the recursion make row and column traversals of the matrix
crucial. Intensive conversion tasks are required for row and column
traversals of the matrices. This makes the overhead of index
conversion essential in the comparison of the different layouts
available. Let $\Theta$ denote a subscript associated with one of the four layouts named above. Given a Cartesian index $(i,j)$, encoding
it in the order $\Theta$ corresponds to calculating its index
$z_{\Theta}$ in the resulting matrix layout under $\Theta$.

\begin{def}
\label{def:decoding}Given an index $z_{\Theta}$ of a matrix entry
under order $\Theta$, decoding $z_{\Theta}$ corresponds to
calculating the Cartesian index $(i,j)$.
\end{def}

\subsection{Conversion Terms}\label{sec:terms}
The following is a list of terminologies used in the remainder of
this manuscript.
\begin{enumerate}
\item{\label{andop} $\&$ : the bitwise AND operator}
\item{\label{orop} $|$ : the bitwise OR operator}
\item{\label{leftshift} $<<$ : the bitwise left shift operator, where $i << k$
is equivalent to multiplying $i$ by $2^{k}$.}
\item{\label{rightshift} $>>$ : the bitwise right shift operator,
where $i >> k$ is equivalent to dividing $i$ by $2^{k}$.}
\item{\label{masks} Masks: These are bit values represented in
hexadecimal format. The masks used hereafter are:
\begin{itemize}
\item{$0$x$00FF00FF = (00000000111111110000000011111111)_{2}$}
\item{$0$x$0F0F0F0F = (00001111000011110000111100001111)_{2}$}
\item{$0$x$33333333 = (00110011001100110011001100110011)_{2}$}
\item{$0$x$55555555 = (01010101010101010101010101010101)_{2}$}
\item{$0$x$0000FFFF = (00000000000000001111111111111111)_{2}$}
\item{$0$x$AAAAAAAA = (10101010101010101010101010101010)_{2}$}
\end{itemize}
}
\item{\label{masking}The act of masking a value $v$: applying an AND
operation on $v$ and a mask to extract part of $v$.}
\item{\label{elt_row_index} The row index of an entry $e$ in a
given matrix is its row offset within the matrix. It is given by $i$
from the Cartesian index $(i,j)$ of $e$. The column index of the
entry is its column offset, given by $j$.}
\item{\label{blk_row_index} Consider a matrix $M$ decomposed into
sub-blocks of size $2^{k} \times 2^{k}$. For any given sub-block,
its row index is its row offset within $M$ and its column index is
its column offset within $M$.}
\item{\label{elt_index}The index of an entry in a given matrix laid out in order $\Theta$ is its offset within
the linear array representing the matrix in the layout $\Theta$. This was denoted earlier by $z_{\Theta}$}
\item{\label{blk_index}The index of a sub-block of size $2^{k} \times 2^{k}$ in a given matrix laid out in order $\Theta$ is the offset of this sub-block within
the linear array of objects representing the matrix in the layout
$\Theta$, where each object is a $2^{k} \times 2^{k}$ sub-block.}
\item{We refer to the index of an entry (or block) in the $\Theta$ order as the $\Theta$
index of this entry (or block). For example, when considering a
matrix $M$ laid out in the row-major order, $\Theta$ refers to the
row-major order and the row-major index of an entry (or block) of
$M$ is the index of this entry (or block) in the row-major layout of
$M$.}
\end{enumerate}

\subsection{Encoding/Decoding in the Row-Major Order}
\label{sec:row_major_conv} 
\begin{figure}[h]
  \centering
     {\includegraphics[width=3in]{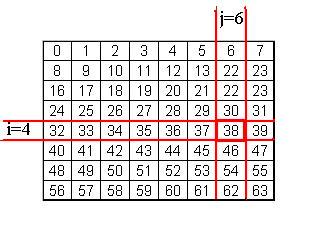}}
  \caption{Row Major Ordered Matrix}
  \label{fig:row_major_indexing}
\end{figure}

The row-major (column-major) order is the default
ordering used by compilers for two dimensional arrays. Computer memory is linear and
consists of a list of consecutive addresses in memory. Compilers therefore 
store a two dimensional array by laying it out row by row. When the programmer uses the notation $A[i][j]$ to access the
element in position $(i,j)$ of the matrix $A$, the compiler performs
the index conversion behind the scenes. In the rest of
this section, we consider the
example matrix shown in Fig. \ref{fig:row_major_indexing} with $n=8$
i.e. $m=3$, and take $\Theta$ to denote the row-major order. 
\subsubsection{Encoding in the Row-Major Order}
Consider the Cartesian index $(i,j)$ in a matrix laid out in a
row-major fashion. To find $z_{\Theta}(i,j,n)$, denoted by
$z_{\Theta}$ for simplicity, use the equation
\begin{equation} \label{eqn:encode_row} z_{\Theta} = i \times n + j.
\end{equation}
%
%
For the example shown in Fig. \ref{fig:row_major_indexing}, to encode
$i=4$ and $j=6$, using Eq. (\ref{eqn:encode_row}) results in
$z_{\Theta}=4 \times 8 + 6 = 32 + 6 = 38$. Because the element of
Cartesian index $(i,j)$ lies within a $2^{m} \times 2^{m}$ matrix,
then $i,j \in \{0, 1, 2, ..., 2^{m}\}$, and so, to represent any of
these values in the binary system, at most $m$ bits are needed.
Write
\begin{equation*}i=(i_{m-1}....i_{3}i_{2}i_{1}i_{0})_{2}
\end{equation*} and
\begin{equation*}j=(j_{m-1}....j_{3}j_{2}j_{1}j_{0})_{2}.\end{equation*}
The integer operations to find $z_{\Theta}$ given by
Eq. (\ref{eqn:encode_row}) are equivalent to the following bit
operations:
\begin{equation*}
z_{\Theta} = ( i << m ) | j
\end{equation*} for $n=2^{m}$. This can be seen as concatenating the bits of $j$ to the bits of
$i$ to get:
\begin{equation*}
z_{\Theta} =
(i_{m-1}...i_{3}i_{2}i_{1}i_{0}j_{m-1}...j_{3}j_{2}j_{1}j_{0})_{2}.
\end{equation*} Hence, the encoding can
be done using bit shifting and bit masking operations on the binary
representations of $i$ and $j$, as shown in
Alg. \ref{alg:encode-row-bit}.
For $i=4$ and $j=6$ represented as $i=(100)_{2}$ and $j=(110)_{2}$,
$i<<3=(100000)_{2}$ and
$z_{\Theta}=(100000)_{2}|(110)_{2}=(100110)_{2}=38$.
\begin{algorithm}

$z_{\Theta}$ = ( $i$ $<<$ $m$ ) $|$ $j$


    \caption{Encoding for the Row-Major Order Using Bit Operations}
  \label{alg:encode-row-bit}
\end{algorithm}

\subsubsection{Decoding in the Row-Major Order}

%
%
\begin{equation}
\label{eqn:decode_row_i} i = z_{\Theta} \div n \end{equation}
\begin{equation}
\label{eqn:decode_row_j} j = z_{\Theta} \% n
\end{equation}

Equations Eq. (\ref{eqn:decode_row_i}) and Eq. (\ref{eqn:decode_row_j})
present the operations used for decoding an index in the row-major
order using integer operations. These equations follow from
Eq. (\ref{eqn:encode_row}) by which we obtain $z_{\Theta}$ as:
\begin{equation*}
z_{\Theta} = i \times n + j
\end{equation*} In Eq. (\ref{eqn:decode_row_i}) and Eq. (\ref{eqn:decode_row_j}), we
extract the $i$ and $j$ value of the index $(i,j)$ respectively.
Using these equations to decode the index 38 from
Fig. \ref{fig:row_major_indexing}, compute $i=38 \div 8 = 4$ and $j =
38 \% 8 = 6$. Alg. \ref{alg:decode_row_bit} describes the
corresponding bit operations.
Recall that $n=2^{m}$, so the extraction of $i$ given by
Eq. (\ref{eqn:decode_row_i}) can be done using the bit operations:
\begin{equation*}
i = z_{\Theta} >> m
\end{equation*} and the extraction of $j$ given by Eq. (\ref{eqn:decode_row_j}) can be done using:
\begin{equation*}
j = z_{\Theta} \& (2^{m}-1 ).
\end{equation*}To decode $z_{\Theta} = 38$, one operates on the bit
representation of $z_{\Theta} = (100110)_{2}$. Extract $i$ as
\begin{equation*}
i=(100110)_{2} >> m=(100110)_{2} >> 3= (100)_{2} = 4
\end{equation*}
and $j$ as
\begin{equation*}
j=(100110)_{2} \& (2^{m}-1)=(100110)_{2} \& (111)_{2} = (110)_{2}=6.
\end{equation*}

\begin{algorithm}
%
%

$i$ = $z_{\Theta}$ $>>$ $m$
\\
$j$ = $z_{\Theta}$ $\&$  $2^{m}-1$


  \caption{Decoding for the Row-Major Order Using Bit Operations}
  \label{alg:decode_row_bit}
\end{algorithm}

\subsubsection{Computation Overhead}
The encoding procedure given by Eq. (\ref{eqn:encode_row}) uses one
integer multiplication and one integer addition and costs two
integer operations. The decoding procedures to extract $i$ and $j$
in Eq. (\ref{eqn:decode_row_i}) and Eq. (\ref{eqn:decode_row_j}) use one
division operation and one modulo operation also costing two integer
operations in total. Equivalently, to encode a Cartesian index in
the row-major order two bit operations are required. Decoding
row-major indices also requires two bit operations.

\subsection{Encoding/Decoding in the Hilbert-based layout}

\begin{figure}[h]
  \centering
     {\includegraphics[width=5in]{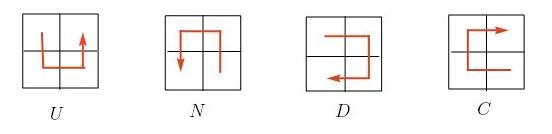}}
  \caption{Primitive Hilbert Patterns}
  \label{fig:hilbert_patterns}
\end{figure}

The Hilbert ordering is recursively generated by dividing a $2^{m} \times 2^{m}$ matrix $M$ into four quadrants following a certain pattern and recursively laying out the elements of these four quadrants \cite{Chatterjee:Layouts}. Before deriving the costs associated with the conversion to and from this layout, We review the procedure to generate the Hilbert order of a $2^{m} \times 2^{m}$ matrix $M$, i.e. to map the entries of $M$ to entries in the one-dimensional array representing $M$. The primitive patterns $U,D,C,$ and $N$ are shown in Fig. \ref{fig:hilbert_patterns}.

In one variant of the Hilbert layout, the matrix $M$ is assigned an
initial pattern $\rho_{M}$ from the four primitive patterns $U,D,C,$
and $N$ and its four quadrants are laid out in an order depending on
the pattern $\rho_{M}$ of $M$. To generate the Hilbert order of $M$,
a recursive procedure is followed. We denote by $Q^{k}$ the
sub-matrix we are laying onto memory for each recursive level and
denote by $\rho_{Q^{k}}$ the pattern of the sub-matrix $Q^{k}$
obtained from the set $\{U,D,C,N\}$. We start with $Q^{0} = M$,
assuming it has pattern $\rho_{Q^{0}}=U$. The generation proceeds as
follows. At the $k^{th}$ step, each of the sub-matrices $Q^{k}$ is
refined into four quadrants $Q^{k+1} \in \{NW_{Q^{k}}, NE_{Q^{k}},
SW_{Q^{k}}, SE_{Q^{k}}\}$ of $Q^{k}$. These are then laid onto
memory in the Hilbert order according to two rules: the
$NextPattern$ rule and the $HilbertOrder$ rule. Let $M_{\rho}$ denote any matrix $M$ of pattern
$\rho \in \{U,D,C,N\}$. Let $i \in \{0,1,2,3\}$ denote the quadrants
of $M_{\rho}$ where $0,1,2,$ and $3$ refer to the $NW$, $NE$, $SW$,
$SE$ quadrants respectively. The $NextPattern(M_{\rho})$ rule for
any sub-matrix $M$ of pattern $\rho_{M} \in \{U,D,C,N\}$ identifies
the pattern of each quadrant of $M$. The $HilbertOrder(M_{\rho})$
rule identifies the order of precedence in which these quadrants are
mapped onto memory.  The $NextPattern$ rule is given by:
\begin{equation*}
NextPattern(M_{\rho}) = \left(
\begin{array}{cc}
\rho_{0} & \rho_{1}\\
\rho_{2} & \rho_{3}
\end{array}
\right)
\end{equation*}
where $\rho_{i}$ denotes the pattern of the $i^{th}$ quadrant of
$M_{\rho}$. The $HilbertOrder$ rule is given by:
\begin{equation*}
HilbertOrder(M_{\rho}) = \left(
\begin{array}{cc}
v_{0} & v_{1}\\
v_{2} & v_{3}
\end{array}
\right)
\end{equation*} where $v_{i}$ represents the order of precedence of
quadrant $i$ in the physical layout when generating the Hilbert
order. The $NextPattern$ rules for the four patterns as used in the
generation of the Hilbert curve are given by the following:

\begin{equation*}
NextPattern(M_{U}) = \left(
\begin{array}{cc}
D & C\\
U & U
\end{array}
\right) \quad \quad NextPattern(M_{D}) = \left(
\begin{array}{cc}
U & D \\
N & D
\end{array}
\right)
\end{equation*}

\begin{equation*}
NextPattern(M_{C}) = \left(
\begin{array}{cc}
C & U \\
C & N
\end{array}
\right) \quad \quad NextPattern(M_{N}) = \left(
\begin{array}{cc}
N & N\\
D & C
\end{array}
\right)
\end{equation*}

The $HilbertOrder$ rules for each pattern are given by the
following:
\begin{equation*}
HilbertOrder(M_{U}) = \left(
\begin{array}{cc}
0 & 3\\
1 & 2
\end{array}
\right) \quad \quad HilbertOrder(M_{D}) = \left(
\begin{array}{cc}
0 & 1 \\
3 & 2
\end{array}
\right)
\end{equation*}

\begin{equation*}
HilbertOrder(M_{C}) = \left(
\begin{array}{cc}
2 & 3 \\
1 & 0
\end{array}
\right) \quad \quad HilbertOrder(M_{N}) = \left(
\begin{array}{cc}
2 & 1\\
3 & 0
\end{array}
\right)
\end{equation*}

\begin{figure}[h]
  \centering
     {\includegraphics[width=6in]{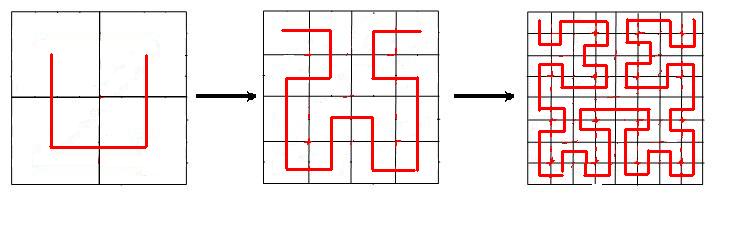}}
  \caption{Generation of Hilbert order}
  \label{fig:hilbert_refinement}
\end{figure}

\begin{figure}[h]
  \centering
     {\includegraphics[width=1.5in]{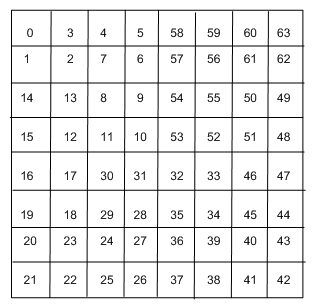}}
  \caption{Matrix in the Hilbert Order}
  \label{fig:hilbert_mat}
\end{figure}

These rules can be deduced from the patterns shown in
Fig. \ref{fig:hilbert_patterns}. In turn, Fig. \ref{fig:hilbert_refinement} shows
the steps of generating the Hilbert order for an $8 \times 8$
matrix $M$ of initial pattern $U$. In Fig. \ref{fig:hilbert_mat}, the
entries of the matrix at the end of the generation of the Hilbert
order show the indices of the elements of $M$ within the physical
one-dimensional array representing $M$ in the Hilbert order.

\label{sec:hilbert_conv} 

 The matrix to be laid out in the Hilbert order is given a pattern and refined according to the refinement rules for
each pattern. The refinement is done by recursively dividing
the matrix into four quadrants and storing them according to the
pattern refinement rules. Let $M_{\rho}$ denote a Hilbert matrix of pattern $\rho \in \{U, D,
C, N\}$. Each step of the refinement for $M_{\rho}$ is done by
identifying the patterns of the quadrants of $M_{\rho}$ and the
order in which these quadrants are accessed. This is done using the
following structures for each matrix $M_{\rho}$ laid out in some
pattern $\rho$:
\begin{equation*}
NextPattern(M_{\rho}) = \left(
\begin{array}{cc}
\rho_{0} & \rho_{1}\\
\rho_{2} & \rho_{3}
\end{array}
\right),
\end{equation*}
where $\rho_{i}$ denotes the pattern of the $i^{th}$ quadrant of
$M_{\rho}$ in the row-major layout, and
\begin{equation*}
HilbertOrder(M_{\rho}) = \left(
\begin{array}{cc}
\nu_{0} & \nu_{1}\\
\nu_{2} & \nu_{3}
\end{array} \right),
\end{equation*}
where $\nu_{i}$ denotes the index in the Hilbert layout of the
$i^{th}$ quadrant of $M_{\rho}$ in the row-major layout.

For example, the refinement rule for the $U$ pattern is given by:
\begin{equation*}
NextPattern(U) = \left(
\begin{array}{cc}
D & C\\
U & U
\end{array}
\right)
\end{equation*}
and
\begin{equation*}
HilbertOrder(U) = \left(
\begin{array}{cc}
0 & 3\\
1 & 2
\end{array} \right).
\end{equation*}

\begin{figure}[h]
  \centering
     {\includegraphics[width=3in]{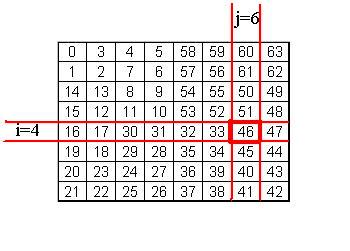}}
  \caption{Hilbert Ordered Matrix}
  \label{fig:hilbert_indexing}
\end{figure}

The refinement process is referred to as the generation of the
Hilbert curve and is going to guide the encoding and decoding
procedures. Fig. \ref{fig:hilbert_indexing} shows an $8 \times 8$
matrix stored in the $U$-shaped Hilbert order on which the encoding
and decoding procedures will be traced. In the rest of this section,
$\Theta$ refers to the Hilbert order and $M$ refers to a $2^{m}
\times 2^{m}$ matrix in the Hilbert order.

\subsubsection{Encoding in the Hilbert Order}
Given an entry $e$ with Cartesian index $(i,j)$ in the $2^{m}
\times 2^{m}$ matrix $M$, we denote by $Q^{k+1}_{e}$ the
quadrant of $Q^{k}_{e}$ of dimensions $2^{(m-(k+1))} \times 2^{(m-(k+1))}$  and in
which the entry $e$ lies in refinement step $k$, for $k=0,1,..m-1$.
We start with $Q^{0}_{e}=M$. To use the refinement rules, we
translate them into two lookup tables: Table $\mathcal{T_{P}}$ and
Table $\mathcal{T_{V}}$. We index the lookup tables using $\rho$ and
$v$: $\rho$ is the pattern of the matrix we are refining and $v$ is
the index, in the row-major order, of the quadrant $Q^{k+1}_{e}$
within $Q^{k}_{e}$. Table $\mathcal{T_{P}}$ is a table mapping a
pattern $\rho$ and an index $v$ to the next pattern, i.e. the
pattern of the block $Q^{k+1}_{e}$, and is given by
Table \ref{table:next_pattern}. The other look-up table,
$\mathcal{T_{V}}$, maps a pattern $\rho$ and an index $v$ to two
bits of $z_{\Theta}$ and is given by Table \ref{table:encode_bits}.
These two bits are the Hilbert index of $Q^{k+1}_{e}$ within
$Q^{k}_{e}$.

To determine these tables, consider matrix $M_{\rho}$ of pattern
$\rho \in \{U, D, C, N\}$. Let $v \in \{0,1,2,3\}$ refer to the
row-major index of any quadrant of $M$, designating $NW$, $NE$, $SW$, and $SE$ respectively. The entry
\begin{equation*}\mathcal{T_{P}}(\rho,v) = \rho^{\prime}\end{equation*}
is determined as follows: $\rho^{\prime}$ is the pattern of the
$v^{th}$ quadrant of $M_{\rho}$ in the row-major order. Table
$\mathcal{T_{V}}$ presents a mapping between the row-major
layout of the quadrants of a matrix of pattern $\rho$ and the
Hilbert layout of these quadrants. The entry
\begin{equation*}\mathcal{T_{V}}(\rho,v) = v^{\prime}\end{equation*}
is the Hilbert index of the $v^{th}$ quadrant of $M_{\rho}$ in the
row-major order.

For example, recall that a matrix of the $U$ Hilbert pattern is refined as follows:
\begin{equation*}
NextPattern(M_{U}) = \left(
\begin{array}{cc}
D & C \\
U & U
\end{array}
\right)
\end{equation*}
and
\begin{equation*}
HilbertOrder(M_{U}) = \left(
\begin{array}{cc}
0 & 3\\
1 & 2
\end{array}
\right).
\end{equation*} Hence the entry $\mathcal{T_{P}}(U,1)$ is
$C$ and $\mathcal{T_{V}}(U,1)=3=(11)_{2}$.

Recall that $Q^{k}_{e}$ denotes the $2^{(m-k)} \times 2^{(m-k)}$
quadrant of $Q^{k-1}_{e}$ in which the element $e$ of Cartesian
index $(i,j)$ lies at each refinement step. Table $\mathcal{T_{P}}$
guides the generation of $z_{\Theta}$ by identifying the pattern of
$Q^{k+1}_{e}$. Table $\mathcal{T_{V}}$ identifies two bits to be
appended at each iteration to a binary index $z_{k}$. These two bits
represent the index, in the Hilbert layout, of $Q^{k+1}_{e}$ within
$Q^{k}_{e}$. We justify that this index is identified using two bits
as follows. Recall that $Q^{k+1}_{e}$ is one of the quadrants of
$Q^{k}_{e}$. There are only four quadrants in a matrix. Hence, the
possible values for the index of any of these quadrants - regardless
of the layout - are 0, 1, 2, and 3, which can be represented using
at most two bits. Hence, the length of the binary representation of
this index is at most two.

\begin{table}
  \centering
  \caption{Encoding Pattern Look-Up Table $\mathcal{T_{P}}$}
  \label{table:next_pattern}
\begin{tabular}{|c|c|c|c|c|}
\hline \textbf{ } & \textbf{0}   & \textbf{1}    & \textbf{2}    &\textbf{3}\\
\hline \textbf{U}     & $D$          &  $C$          & $U$           & $U$\\
\hline \textbf{C}     & $C$          &  $U$          & $C$           & $N$\\
\hline \textbf{D}     & $U$          &  $D$          & $N$           & $D$\\
\hline \textbf{N}     & $N$          &  $N$          & $D$           & $C$\\
\hline
\end{tabular}
\end{table}
\begin{table}
  \centering
  \caption{Encoding Bits Look-Up Table $\mathcal{T_{V}}$}
  \label{table:encode_bits}
\begin{tabular}{|c|c|c|c|c|}
\hline \textbf{ } & \textbf{0}   & \textbf{1}    & \textbf{2}    &\textbf{3}\\
\hline \textbf{U}       & 00           &  11           & 01            & 10\\
\hline \textbf{C}       & 10           &  11           & 01            & 00\\
\hline \textbf{D}       & 00           &  01           & 11            & 10\\
\hline \textbf{N}       & 10           &  01           & 11            & 00\\
\hline
\end{tabular}
\end{table}
Now that the lookup tables are ready, we describe the iterative encoding
procedure for a given $2^{m} \times 2^{m}$ matrix and a Cartesian index
$(i,j)$. Each iteration
represents a refinement step within the generation of the Hilbert
curve and, after $m$ iterations, the sequence $\{z_{k+1}\}_{k = 0,1,...,m-1}$, converges to $z_{\Theta}$. In each iteration
$k$, the pattern $\rho_{k}$ - whether $U$, $C$, $D$, or $N$ - of the
quadrant $Q^{k+1}_{e}$ of $Q^{k}_{e}$ must be determined. Recall
that $Q^{k}_{e}$ is the $2^{(m-k)} \times 2^{(m-k)}$ sub-block of
$M$ containing element $e$ of Cartesian index $(i,j)$. The initial
$Q^{0}_{e}$ is $M$ and $\rho_{0} = U$ because the Hilbert pattern we
assume for the initial matrix is $U$. The corresponding index in the
Hilbert order is progressively calculated by finding some partial
index $z_{k}$ in each iteration. The algorithm begins with $z_{0} =
0$ and ends with the value for $z_{m}=z_{\Theta}$. Write
\begin{equation*}
i = (i_{m-1}...i_{2}i_{1}i_{0})_{2}
\end{equation*} and
\begin{equation*}
j = (j_{m-1}...j_{2}j_{1}j_{0})_{2}.
\end{equation*}Recall that there are $m$ bits in each of $i$ and $j$,
because at most $m$ bits are needed to represent a value between 0
and $2^{m-1}$ in the binary system. In each iteration $k$ of the encoding algorithm:
\begin{itemize}
\item{we generate an auxiliary index $v_{k} = (i_{(m-1-k)}j_{(m-1-k)})_{2}$, where $v_{k} \in \{0,1,2,3\}$, the two-bit, row-major
index of $Q^{k+1}_{e}$ within $Q^{k}_{e}$.}
\item{we use $v_{k}$ and $\rho_{k}$ to index the lookup table $\mathcal{T_{P}}$ and find the next pattern of $Q^{k+1}_{e}$ $\rho_{k+1} = \mathcal{T_{P}}(\rho_{k},v_{k})$.}
\item{we use $v_{k}$ and $\rho_{k}$ to index the lookup table $\mathcal{T_{V}}$ and
find
\begin{equation*}z_{k+1} = (z_{k} << 2) | \mathcal{T_{V}}(\rho_{k},v_{k}).\end{equation*} This appends the two bits of the Hilbert index of $Q^{k+1}_{e}$ within $Q^{k}_{e}$ to the partial index $z_{k}$ to get $z_{k+1}$.}
\end{itemize}

As two bits are appended to $z_{k}$ at each iteration, and there are
$m$ iterations, then $z_{\Theta}$ is formed of $2m$ bits.

To prove correctness, we propose the following:
\begin{lem}\label{prps:row_ind} The auxiliary variable $v_{k}$ is the row-major index of $Q^{k+1}_{e}$ within
$Q^{k}_{e}$. \end{lem}

\paragraph{proof} Consider the entry $e$ of Cartesian index $(i,j)$. Write
\begin{equation*}
i = (i_{m-1}...i_{2}i_{1}i_{0})_{2}
\end{equation*} and
\begin{equation*}
j = (j_{m-1}...j_{2}j_{1}j_{0})_{2}.
\end{equation*} To prove that $v_{k}$ is the row-major index of $Q^{k+1}_{e}$
within $Q^{k}_{e}$, we will show that the bits $i_{(m-1-k)}$ and
$j_{(m-1-k)}$ are the row and column indices of the
$Q^{k+1}_{e}$ within $Q^{k}_{e}$ respectively, from which it follows
directly that $v_{k}$, the concatenation of $j_{(m-1-k)}$ to
$i_{(m-i-k)}$, is the row-major index of $Q^{k+1}_{e}$ within
$Q^{k}_{e}$. We proceed by induction on $k$.

\noindent \textbf{Base Case:} For $k=0$, recall that $Q^{0}_{e}$ is
the $2^{m} \times 2^{m}$ matrix $M$. Let $i_{B}$ denote the row
index of $Q^{1}_{e}$ within $Q^{0}_{e}$. We need to show that
$i_{B}=i_{(m-1)}$. Let $i_{r}$ denote the row index of $e$ within
$Q^{1}_{e}$. The row index $i$ of $e$ is given by $i = i_{B} \times
2^{m-1} + i_{r}$ since $Q^{1}_{e}$ is of size $2^{m-1} \times
2^{m-1}$. The equation $i = i_{B} \times 2^{m-1} + i_{r}$ is
equivalent to
\begin{equation*}
i = ( i_{B} << (m-1) ) | i_{r}
\end{equation*}
in bit operations . Hence, $i_{B}$ is the $(m-1)^{st}$ bit of $i$
given by $i_{(m-1)}$. This establishes the base case.

\noindent \textbf{Induction Step:}  We want to show that
$i_{(m-(k+1))}$ is the row index of $Q^{k+1}_{e}$ within
$Q^{k}_{e}$. Let $i_{B}$ denote the row index of $Q^{k+1}_{e}$
within $Q^{k}_{e}$. We need to show that $i_{B} = i_{(m-(k+1))}$.
Let $i_{r}$ and $i^{\prime}_{r}$ denote the row index of $e$ within
$Q^{k}_{e}$ and $Q^{k+1}_{e}$ respectively. As $Q^{k+1}_{e}$ is a
$2^{m-(k+1)} \times 2^{m-(k+1)}$ quadrant within $Q^{k}_{e}$, we
have
\begin{equation*}
i_{r} = i_{B} \times 2^{m-(k+1)} + i'_{r}
\end{equation*} which is equivalent to
\begin{equation}
\label{eqn:one} i_{r} = ( i_{B} << (m-(k+1)) ) | i'_{r}.
\end{equation}
By induction, we know that the bit $i_{(m-1-k)}$ is the row index of $Q^{k}_{e}$ within $Q^{k-1}_{e}$,
so that
\begin{equation}
\label{eqn:two} i = (i_{m-1}i_{m-2}...i_{(m-k)} << (m-k) ) | i_{r}.
\end{equation} Replacing $i_{r}$ as in Eq. (\ref{eqn:one}) in Eq. (\ref{eqn:two}) above, we obtain
\begin{equation*}
i = (i_{m-1}i_{m-2}...i_{(m-k)} << (m-k) ) | ( i_{B} << (m-(k+1)) )
| i'_{r}.
\end{equation*} Hence the $i_{(m-(k+1))} = _{B}$. This establishes the induction step and concludes the
proof.

\begin{prop}
The partial index $z_{k}$ is the Hilbert index of the sub-block $Q^{k}_{e}$ of dimenions $2^{(m-k)}
\times 2^{(m-k)}$ within $M$, for
$k=0,1,...m$.
\end{prop}

\paragraph{proof} We proceed by induction on $k$.

\noindent \textbf{Base Case:} For $k=0$, $Q^{k}_{e} \equiv M$ and has
index $z_{k} = z_{0} = 0$ within the Hilbert layout of $M$.

\noindent \textbf{Induction Step:} We assume that $z_{k}$ is the
Hilbert index of the $2^{(m-k)} \times 2^{(m-k)}$ sub-block
$Q^{k}_{e}$ within $M$. We refine more and obtain the $2^{(m-(k+1))}
\times 2^{(m-(k+1))}$ quadrant $Q^{k+1}_{e}$ within $Q^{k}_{e}$.
From Prop. \ref{prps:row_ind}, we know that $v_{k}$ is the row-major
index of $Q^{k+1}_{e}$ within $Q^{k}_{e}$. The quadrant
$Q^{k+1}_{e}$ has Hilbert index $v^{\prime}_{k} =
\mathcal{T_{V}}(\rho_{k},v_{k})$ within $Q^{k}_{e}$ from the
definition of table $\mathcal{T_{V}}$. We will show that
\begin{equation*}
z_{k+1} = ( z_{k} << 2 ) | v^{\prime}_{k}
\end{equation*}
is the Hilbert index of $Q^{k+1}_{e}$ within $M$. By induction,
$z_{k}$ represents the Hilbert index of $Q^{k}_{e}$ within $M$. Upon
another refinement step, each of the $2^{(m-k)} \times 2^{(m-k)}$
sub-blocks of $M$ is in turn divided into four quadrants of size
$2^{(m-(k+1))} \times 2^{(m-(k+1))}$ each. The indices of the four
quadrants of $Q^{k}_{e}$ are obtained
by $z_{k} \times 4$, $z_{k} \times 4 + 1$, $z_{k} \times 4 + 2$, and
$z_{k} \times 4 + 3$ depending on their index $v^{\prime}_{k}$ in
the Hilbert layout of $Q^{k}_{e}$. These indices can be re-written
as $(z_{k} << 2) | 00$, $(z_{k} << 2) | 01$, $(z_{k} << 2) | 10$,
and $(z_{k} << 2) | 11$. The sub-block $Q^{k+1}_{e}$ is a quadrant
of $Q^{k}_{e}$ and thus $Q^{k+1}_{e}$ takes on one of these indices
depending on its Hilbert index $v^{\prime}_{k}$ within $Q^{k}_{e}$.
Thus $Q^{k+1}_{e}$ has index
\begin{equation*}
z_{k+1} = (z_{k} << 2) | v^{\prime}_{k}
\end{equation*} within $M$. This concludes the proof.

\begin{cor} The index $z_{m}$ is $z_{\Theta}$.
\end{cor}

\paragraph{proof}
For $k=m$, $z_{m}$ is the Hilbert index of the smallest sub-block
$Q^{m}_{e}$ within $M$ containing $e$, which is $e$ itself. This
concludes the proof.

We illustrate this encoding procedure with an example in Appendix \ref{A1}.

\subsubsection{Decoding in the Hilbert Order}
In the decoding procedure, we are given the Hilbert index
$z_{\Theta}$ of an entry $e$ and we wish to find $(i,j)$, the
Cartesian index of $e$. Recall that $Q^{k}_{e}$ denotes the quadrant
of $Q^{k-1}_{e}$ in which $e$ lies, starting with $Q^{0}_{e} = M$.
To find $(i,j)$, we need two different lookup tables:
$\mathcal{T^{\prime}_{P}}$ (Table \ref{table:next_pattern_dec} below) 
is used to identify the pattern
of $Q^{k}_{e}$. Table $\mathcal{T^{\prime}_{V}}$ (Table \ref{table:decode_bits} below) 
is used to find a two-bit value
$v^{\prime}_{k}$ representing the row-major index of $Q^{k}_{e}$
within $Q^{k-1}_{e}$. To determine these tables, consider matrix
$M_{\rho}$ of pattern $\rho \in \{U, D, C, N\}$. Recall that the
$NW$, $NE$, $SW$, and $SE$ quadrants are the $0^{th}$, $1^{st}$,
$2^{nd}$, $3^{rd}$ quadrants in the row-major order. Entries in
Table $\mathcal{T^{\prime}_{P}}$ represent patterns and are given
by:
\begin{equation*}
\mathcal{T^{\prime}_{P}}(\rho,v) = \rho^{\prime}.
\end{equation*} The pattern $\rho^{\prime}$ refers to the pattern
of the $v^{th}$ quadrant of $M_{\rho}$ in the Hilbert order. Table 
$\mathcal{T^{\prime}_{V}}$ presents a mapping between the
Hilbert layout of the quadrants of $M_{\rho}$ and the row-major
layout of these quadrants. The entry
\begin{equation*}\mathcal{T^{\prime}_{V}}(\rho,v) = v^{\prime}\end{equation*}
indicates that the $v^{th}$ quadrant of $M_{\rho}$ in the Hilbert order
is the $v^{\prime th}$ quadrant of $M_{\rho}$ in the row-major
order. Note that $\mathcal{T^{\prime}_{V}}$ is the inverse mapping
of the table $\mathcal{T_{V}}$ used for encoding: i.e.
\begin{equation*}\mathcal{T^{\prime}_{V}}(\rho,v) = v^{\prime} \leftrightarrow
\mathcal{T_{V}}(\rho, v^{\prime}) = v.\end{equation*}

For example, recall that a matrix $M_{U}$ in a $U$ Hilbert pattern
is refined as follows:
\begin{equation*}
NextPattern(M_{U}) = \left(
\begin{array}{cc}
D & C \\
U & U
\end{array}
\right),
\end{equation*} and
\begin{equation*}
HilbertOrder(M_{U}) = \left(
\begin{array}{cc}
0 & 3 \\
1 & 2
\end{array}
\right).
\end{equation*}
Hence, we have $\mathcal{T^{\prime}_{P}}(U,3)=C$ because the
$3^{rd}$ quadrant of $M_{U}$ in the Hilbert order has pattern $C$.
Also, $\mathcal{T^{\prime}_{V}}(U,3)=1=(01)_{2}$ because the
$3^{rd}$ quadrant of $M_{U}$ in the Hilbert order is the $1^{st}$ -
i.e. the $NE$ quadrant of $M_{U}$ in the row-major order. Tables
$\mathcal{T^{\prime}_{P}}$ and $\mathcal{T^{\prime}_{V}}$ guide the
generation of $i$ and $j$ by identifying the pattern of $Q^{k}_{e}$
and the row-major index of $Q^{k}_{e}$ within $M$ respectively.

\begin{table}
  \centering
  \caption{Decoding Pattern Look-Up Table $\mathcal{T^{\prime}_{P}}$}
  \label{table:next_pattern_dec}
\begin{tabular}{|c|c|c|c|c|}
\hline \textbf{ } & \textbf{0}   & \textbf{1}    & \textbf{2}    &\textbf{3}\\
\hline \textbf{U}     & $D$          &  $U$          & $U$           & $C$\\
\hline \textbf{C}     & $N$          &  $C$          & $C$           & $U$\\
\hline \textbf{D}     & $U$          &  $D$          & $D$           & $N$\\
\hline \textbf{N}     & $C$          &  $N$          & $N$           & $D$\\
\hline
\end{tabular}
\end{table}

\begin{table}
  \centering
  \caption{Decoding Bits Look-Up Table $\mathcal{T^{\prime}_{V}}$}
  \label{table:decode_bits}
\begin{tabular}{|c|c|c|c|c|}
\hline \textbf{ } & \textbf{0}   & \textbf{1}    & \textbf{2}    &\textbf{3}\\
\hline \textbf{U}       & 00           &  10           & 11            & 01\\
\hline \textbf{C}       & 11           &  10           & 00            & 01\\
\hline \textbf{D}       & 00           &  01           & 10            & 11\\
\hline \textbf{N}       & 11           &  01           & 00            & 10\\
\hline
\end{tabular}
\end{table}

Now that the lookup tables have been described, we present the
decoding procedure. Recall from the encoding procedure that
$z_{\Theta}$ is formed of $2m$ bits. Write
\begin{equation*}
z_{\Theta} = (z_{2m-1}...z_{2}z_{1}z_{0})_{2}. \end{equation*} The
decoding reverses the encoding procedure, in which at each iteration
one bit of each of $i$ and $j$ is used to generate two bits of $z$.
Recall, from encoding, that, in each iteration $k$, these two bits
of $z_{\Theta}$ make up the Hilbert index of $Q^{k}_{e}$ within
$Q^{k-1}_{e}$ as follows:
\begin{equation*}
v_{k}=(z_{(2m-1-2k)}z_{(2m-1-(2k+1))})_{2}. \end{equation*} Also, we
identify the pattern $\rho_{k}$ of $Q^{k}_{e}$. We start with
$\rho_{0} = U$. In each iteration $k$, $\rho_{k}$ and $v_{k}$ are
used to index the two lookup tables $\mathcal{T^{\prime}_{P}}$ and
$\mathcal{T^{\prime}_{V}}$. As mentioned earlier, table
$\mathcal{T^{\prime}_{P}}$ is used to find the pattern of the
quadrant of the next iteration $Q^{k+1}_{e}$, given by $\rho_{k+1} =
\mathcal{T^{\prime}_{P}}(\rho_{k},v_{k})$. Table
$\mathcal{T^{\prime}_{V}}$ is used to find a two-bit value
$v^{\prime}_{k} = (b_{i}b_{j})_{2}$ which represents the row-major
index of $Q^{k+1}_{e}$ within $Q^{k}_{e}$. Recall that this is a
two-bit value because it represents the index of a quadrant. The
values of $i$ and $j$ are found progressively: start with $i_{0}=0$
and $j_{0}=0$. In each iteration append $b_{i} \in \{0,1\}$ and
$b_{j} \in \{0,1\}$ to $i_{k}$ and $j_{k}$ to get $i_{k+1}$ and
$j_{k+1}$ respectively. In bit operations, this corresponds to
\begin{equation*} i_{k+1} = ( i_{k} << 1 ) | (v^{\prime}_{k}
>> 1 ),\end{equation*} 
appends bit $b_{i}$ to $i_{k}$ to get $i_{k+1}$, and
\begin{equation*} j_{k+1} = (j_{k} << 1 ) | (v^{\prime}_{k} \& 1),\end{equation*}
which appends bit $b_{j}$ to $j_{k}$ to get
$j_{k+1}$. This reverses the process of encoding. We iterate $m$
times, after which we have $i_{m}=i$ and $j_{m}=j$.

We illustrate this decoding procedure with an example in Appendix \ref{A2}.

\subsubsection{Computation Overhead}

Each of encoding and decoding requires $m$ iterations. In each
iteration, the encoding operation uses six bit operations and two
table look-ups and the decoding algorithm uses eight bit operations
and two table look-ups. The first of each table look-up incurs a random cache miss. The tables are 
small enough to fit in internal memory. As row and column permutations swaps in TURBO take place consecutively in one batch, so do the conversion routines, each of which requires access to the look-up table. By the LRU cache policy, the tables are kept in internal memory until the permutations have concluded. Consequently, the overall I/O cost for table lookups is $\Theta(1)$ per one batch of permutations, which is dominated by the I/O cost of the matrix operations in each recursive step, and hence, can be discarded. Summarising, the overall run-time for each of encoding and decoding in the Hilbert order is $\Theta(m)$ bit operations. 

Algorithms for encoding and decoding in the Hilbert order which do not use look-up tables are described in full detail in \cite{Hilbert:EncDec}. In \cite{Hilbert:EncDecNew}, a new variant of the Hilbert curve is introduced for which the encoding runtime overhead is $O(\lg(\max(i,j)))$. Yet, such conversion schemes are still beaten by the Morton and Morton-hybrid layouts as we will describe in Sec. \ref{sec:morton_conv} and Sec. \ref{sec:morton_hybrid_conv} respectively.

\subsection{Encoding/Decoding in the Peano Layout}

\begin{figure}[h]
  \centering
     {\includegraphics[width=1.5in]{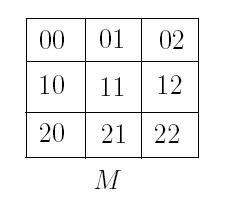}}
  \caption{Peano Matrix Division}
  \label{fig:peano_division}
\end{figure}

The Peano ordering of a matrix is based on the Peano space filling
curve. This ordering results from recursive construction and assumes
the matrix has dimension $3^{m} \times 3^{m}$ \cite{Peano:Mult}. Each
dimension of $M$ is divided into $3$ equal parts as shown in
Fig. \ref{fig:peano_division} resulting in nine sub-matrices which are
named $\{00,01,02,10,11,12,20,21,22\}$. The Peano order for
each of the resulting sub-matrices is recursively generated
according to a certain precedence depending on the pattern of the
higher level sub-matrix.
\begin{figure}[h]
  \centering
     {\includegraphics[width=6in]{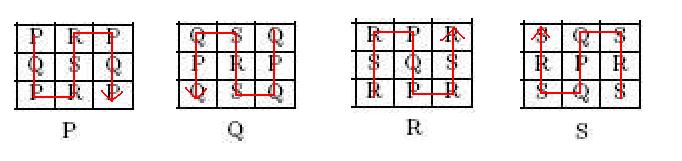}}
  \caption{Peano Generation Rules}
  \label{fig:peano_rules}
\end{figure}
Any Peano sub-matrix has a pattern from the set of primitive
patterns $\{P,Q,R,S\}$ shown in Fig. \ref{fig:peano_rules}. The Peano
order assumes the initial matrix $M$ has pattern $P$. We denote by
$Q^{k}$ the sub-matrix we are laying onto memory for each recursive
level and denote by $\rho_{Q^{k}}$ the pattern of the sub-matrix
$Q^{k}$ obtained from the set $\{P,Q,R,S\}$. We start with $Q^{0} =
M$ of pattern $\rho_{Q^{0}}=P$. At the $k^{th}$ step, each of the sub-matrices $Q^{k}$ is refined
into nine quadrants $Q^{k+1} \in \{00_{Q^{k}}, 01_{Q^{k}},
02_{Q^{k}}, 10_{Q^{k}}, 11_{Q^{k}}, 12_{Q^{k}}, 20_{Q^{k}},
21_{Q^{k}}, 22_{Q^{k}}\}$ of $Q^{k}$. These are then laid onto
memory in the Peano order governed by two rules: the $NextPattern$
rule and the $PeanoOrder$ rule, similar to the rules governing the
generation of the Hilbert order. Let $M_{\rho}$ denote any matrix
$M$ of pattern $\rho \in \{P,Q,R,S\}$. Let $i \in
\{0,1,2,3,4,5,6,7,8\}$ denote the sub-matrices of $M_{\rho}$ where
each $i$ refers to the $00$, $01$, $02$, $10$, $11$,
$12$, $20$, $21$, and $22$ sub-matrices respectively. The
$NextPattern(M_{\rho})$ rule for any sub-matrix $M$ of pattern
$\rho_{M} \in \{P,Q,R,S\}$ identifies the pattern of each quadrant
of $M$. The $PeanoOrder(M_{\rho})$ rule identifies the order of
precedence in which these quadrants are mapped onto memory. These
rules can also be seen from Fig. \ref{fig:peano_rules}: the precedence
order of the sub-matrices is given by the direction of the arrow and
the pattern is given by the letter inside the sub-matrix.

\begin{figure}[h]
  \centering
     {\includegraphics[width=5in]{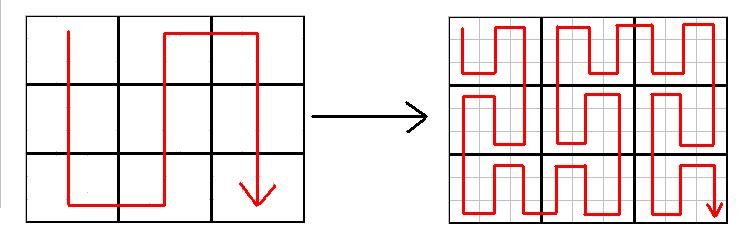}}
  \caption{Generation of Peano order}
  \label{fig:peano_refinement}
\end{figure}

\begin{figure}[h]
  \centering
     {\includegraphics[width=2in]{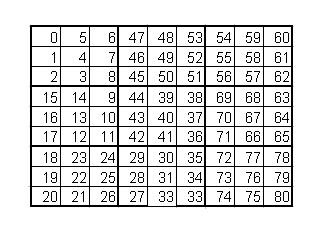}}
  \caption{Matrix in the Peano Order}
  \label{fig:peano_mat}
\end{figure}

Fig. \ref{fig:peano_refinement} shows the steps for the generation of
the Peano order for a $9 \times 9$ matrix $M$. In
Fig. \ref{fig:peano_mat}, the entries of the matrix at the end of the
generation of the Peano order show the indices of the elements of
$M$ within the physical one-dimensional array representing $M$ in
the Peano order.

The Peano-based ordering of matrices has the property that computations within 
matrix-matrix multiplication can be re-ordered so
as to ensure no jumps in the address space \cite{Peano:Mult}.

The Peano layout is similar to the Hilbert layout in that it is based on more than one pattern, and so
the encoding and decoding procedures for the Peano layout are
analogous to those for the Hilbert layout. Specifically, the indices
are encoded (or decoded) progressively through iterations. In each
iteration, a pattern identifier is used along with a set of lookup
tables to identify auxiliary variables for the next iteration. The
differences between the procedure for the Peano layout and those for
the Hilbert layout are:
\begin{itemize}
\item{The operations used in each iteration are operations in base 3 - division by 3 and modulo 3.}
\item{The resulting indices are in base 3 and need to be converted to base 2.}
\end{itemize}
This results in $m$ iterations with a constant number of operations
in base 3 per iteration, which cannot be replaced by bit
operations.
Conversion between base 3 and base 2 values is also needed at the end of the iterative computations in order to get the final indices. In total, this makes encoding and decoding indices within the Peano layout significantly costly as opposed to the Hilbert, Morton, and Morton-hybrid orders. As such, we rule out using the Peano order within the TURBO algorithm. This is specifically so because the nature of the recursive TU decomposition algorithms requires row-wise and column-wise traversal of the matrix for pivoting and permutations -- in contrast to matrix multiplication where any traversal that suits the layout may be used. 

\subsection{Encoding/Decoding in the Morton Layout}
\label{sec:morton_conv} 

\begin{figure}[h]
  \centering
     {\includegraphics[width=1.5in]{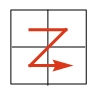}}
  \caption{Morton $Z$ Pattern}
  \label{fig:morton_patterns}
\end{figure}

The Morton order is another alternative layout used for $2^{m} \times 2^{m}$ matrices. To generate the
Morton-order of a matrix, the latter is divided into four quadrants, which
are laid out onto memory in the order northwest, northeast,
southwest, then southeast. The Morton order differs from the Hilbert
order in that it does not alternate between patterns, but rather
uses the same pattern to generate the Morton order for each of the
sub-matrices. In this section, we present the $Z$-shaped Morton
order, in which the pattern used looks like the letter Z as can be
seen in Fig. \ref{fig:morton_patterns}.

\begin{figure}[h]
  \centering
     {\includegraphics[width=6in]{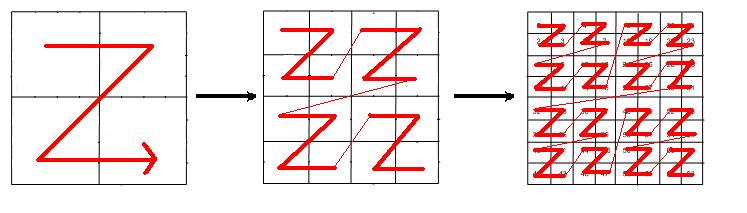}}
  \caption{Generation of Morton order}
  \label{fig:morton_refinement}
\end{figure}

\begin{figure}[h]
  \centering
     {\includegraphics[width=1.5in]{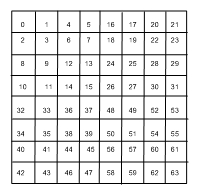}}
  \caption{Matrix in Morton Order}
  \label{fig:morton_mat}
\end{figure}

Fig. \ref{fig:morton_refinement} shows the steps in the generation of
the $Z$-shaped Morton order of an $8 \times 8$ matrix. For this
example, there are three levels of decomposition, after which the
entries of the matrix are laid out as shown in
Fig. \ref{fig:morton_mat}.

\begin{figure}[h]
  \centering
     {\includegraphics[width=3in]{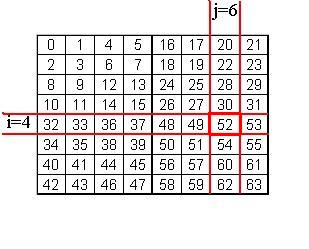}}
  \caption{Morton Ordered Matrix}
  \label{fig:morton_indexing}
\end{figure}

In the remainder of this section,
$\Theta$ refers to the Morton layout and we assume a given matrix
$M$ of dimensions $2^{m} \times 2^{m}$. Given a Cartesian index
$(i,j)$, recall that the length of the binary representation of $i$
and $j$ is at most $m$ because $i,j \in \{0,1,2,...,2^{m}-1\}$.
These values can be represented by at most $m$ bits. Write
\begin{equation*}i = (i_{m-1}...i_{4}i_{3}i_{2}i_{1}i_{0})_{2}\end{equation*} and
\begin{equation*}j=(j_{m-1}...j_{4}j_{3}j_{2}j_{1}j_{0})_{2}.\end{equation*}
According to \cite{Morton:EncDec}, the corresponding Morton index
$z_{\Theta}$ is
\begin{equation}\label{eqn:interleaving}
z_{\Theta}=(i_{m-1}j_{m-1}...i_{4}j_{4}i_{3}j_{3}i_{2}j_{2}i_{1}j_{1}i_{0}j_{0})_{2},
\end{equation}
which represents an inter-leaving of the bits of $i$ and $j$. The
reverse process by which we extract the bits of $i$ and $j$ from the
bits of $z_{\Theta}$ is referred to as de-leaving. In this section,
we derive the costs for encoding and decoding the Morton order
based on this inter-leaving of indices. For this, we revisit the
notions of dilating and un-dilating integers from \cite{Morton:EncDec}.
\begin{def}\label{def:dilation}Dilating an integer $k =
(k_{b-1}...k_{4}k_{3}k_{2}k_{1}k_{0})_{2}$ of $b$ bits is the
process of expanding $k$ into an integer $k^{\prime} =
(0k_{b-1}...0k_{4}0k_{3}0k_{2}0k_{1}0k_{0})_{2}$ of $2 \times b$
bits by inserting a zero bit between every two bits of $k$.
Un-dilating is the reverse process that takes $k^{\prime}$ back to
$k$.
\end{def}
Before we determine the computational overhead for index conversion, we review here two algorithms - \textit{dilate} and
\textit{un-dilate} - that perform the encoding and decoding between Morton order indices and Cartesian
indices. Fig. \ref{fig:morton_indexing} shows an $8 \times 8$ Morton
ordered matrix on which the encoding and decoding procedures will be
traced.

\subsubsection{Encoding in the Morton Order}

\begin{algorithm}
%
%
unsigned int $r$ = $t$\\
$r$ = ($r$ $|$ ($r$ $<<$ 8)) \& $0$x$00FF00FF$ \\
$r$ = ($r$ $|$ ($r$ $<<$ 4)) \& $0$x$0F0F0F0F$ \\
$r$ = ($r$ $|$ ($r$ $<<$ 2)) \& $0$x$33333333$ \\
$r$ = ($r$ $|$ ($r$ $<<$ 1)) \& $0$x$55555555$ \\
return $r$


  \caption{(unsigned int) \textit{dilate}( unsigned short t) \cite{DilatedInts}}
  \label{alg:dilate_morton}
\end{algorithm}

All the present discussion follows from \cite{Morton:EncDec,DilatedInts}. For encoding the Morton order, $z_{\Theta}$ must be found given $(i,j)$. As shown in Eq. (\ref{eqn:interleaving}), the inter-leaving of the bits of $i$ and $j$ must be performed. This can be done by finding the dilated forms of $i$ and $j$, denoted by $i^{\prime}$ and $j^{\prime}$ respectively. The algorithm for dilation is given by Alg. \ref{alg:dilate_morton} from \cite{DilatedInts}. According to the definition of dilation above, dilating $i$ and $j$ gives 
\begin{equation*}
i^{\prime} = (0i_{m-1}0i_{m-2}...0i_{3}0i_{2}0i_{1}0i_{0})_{2}
\end{equation*}
and \begin{equation*}
j^{\prime}=(0j_{m-1}0j_{m-2}...0j_{3}0j_{2}0j_{1}0j_{0})_{2}.
\end{equation*} Then
$z_{\Theta}$ is given by
\begin{equation}
\label{eqn:calc_morton} z_{\Theta} = ( i^{\prime} << 1 ) |
j^{\prime}. 
\end{equation} To verify Eq. (\ref{eqn:calc_morton}) from \cite{Morton:EncDec}, perform
\begin{equation*}
i^{\prime} << 1 = (i_{m-1}0i_{m-2}0...i_{3}0i_{2}0i_{1}0i_{0}0)_{2}
\end{equation*} followed by
\begin{equation*}
( i^{\prime} << 1 ) | j^{\prime} =
(i_{m-1}j_{m-1}i_{m-2}j_{m-2}...i_{3}j_{3}i_{2}j_{2}i_{1}j_{1}i_{0}j_{0})_{2}.
\end{equation*}

We illustrate this encoding procedure with an example in Appendix \ref{A3}.

\subsubsection{Decoding in the Morton Order}
To decode the Morton order, we need to extract $i$ and $j$ from
$z_{\Theta}$. We write $z_{\Theta}$ as in Eq. (\ref{eqn:interleaving}).
Then, the bits of $z_{\Theta}$ must be de-leaved to separate the
bits of $i$, denoted $i_{z_{\Theta}}$, from the bits of $j$, denoted
$j_{z_{\Theta}}$. To do this, we mask $z_{\Theta}$ with $0$x$AAAAAAAA$
to get
\begin{equation*}
i_{z_{\Theta}}=(i_{m-1}0i_{m-2}0...i_{3}0i_{2}0i_{1}0i_{0}0)_{2}.
\end{equation*} and mask $z_{\Theta}$ with
$0$x$55555555$ to get \begin{equation*}
j_{z_{\Theta}}=(0j_{m-1}0j_{m-2}...0j_{3}0j_{2}0j_{1}0j_{0})_{2}.
\end{equation*}
Now we shift $i_{z_{\Theta}}$ one position to the right to get
\begin{equation*}
i^{\prime}_{z_{\Theta}} = i_{z_{\Theta}} >> 1 =
(0i_{m-1}0i_{m-2}...0i_{3}0i_{2}0i_{1}0i_{0})_{2}.
\end{equation*} To calculate $i$, we un-dilate $i^{\prime}_{z_{\Theta}}$ using
the un-dilation algorithm given in Alg. \ref{alg:undilate_morton}. To
calculate $j$, we un-dilate $j_{z_{\Theta}}$ using
Alg. \ref{alg:undilate_morton}.

\begin{algorithm}
%
 %
unsigned int $r$ = $t$\\
$t$ = ($t$ $|$ ($t$ $>>$ 1)) \& $0$x$33333333$ \\
$t$ = ($t$ $|$ ($t$ $>>$ 2)) \& $0$x$0F0F0F0F$ \\
$t$ = ($t$ $|$ ($t$ $>>$ 4)) \& $0$x$00FF00FF$ \\
$t$ = ($t$ $|$ ($t$ $>>$ 8)) \& $0$x$0000FFFF$ \\
return (unsigned short)$t$


  \caption{unsigned short \textit{un-dilate}( unsigned int t) taken from \cite{Morton:EncDec,DilatedInts}}
  \label{alg:undilate_morton}
\end{algorithm}
We illustrate this decoding procedure with an example in Appendix \ref{A4}.

\subsubsection{Computation Overhead}
Assume that the input matrix has dimensions $2^{\alpha} \times 2^{\alpha}$, 
where $\alpha$ is the machine word-size. This guarantees that each cartesian index fits in a machine word. For the typical value $\alpha = 64$, such matrix sizes are very generous. Each of dilation and un-dilation performs twelve bit operations: four OR operations, four AND operations, and four bit-shift operations. The encoding procedure makes use of the dilation algorithm twice, followed by one bit-shift operation and one OR operation. This results in a total of twenty six bit operations for encoding an index in the Morton order. The decoding procedure makes use of two masking operations, one bit-shift operation, and two calls to the un-dilation algorithm: one for un-dilating $i_{z}$ and one for un-dilating $j_{z}$. This requires a total of twenty seven bit operations for decoding a Morton index to a Cartesian index.

\begin{figure}[h]
  \centering
     {\includegraphics[width=6in]{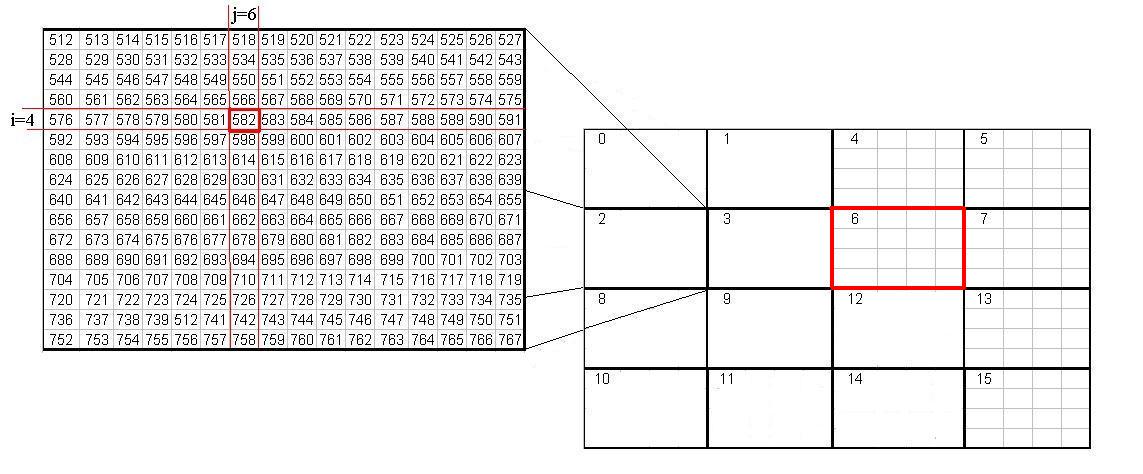}}
  \caption{Morton-Hybrid Ordered Matrix with $T=16$}
  \label{fig:morton_hybrid_indexing1}
\end{figure}

\subsection{Encoding/Decoding in the Morton-Hybrid Layout}
\label{sec:morton_hybrid_conv}

The Morton-hybrid order is a variant of the Morton order used in \cite{Wise:MaskedIntegers} and \cite{Chatterjee:Layouts} among others. In this variant of the Morton order, the recursive sub-dividing into blocks stops at a given truncation size $T$, resulting in a base case block of size $T \times T$, for which the row-major order is used. The value of the truncation size can vary.

\begin{figure}[h]
  \centering
     {\includegraphics[width=6in]{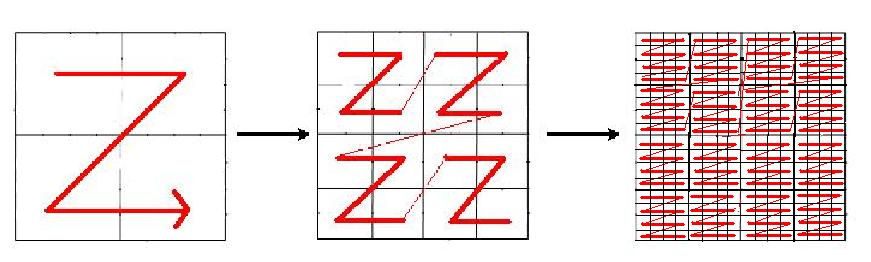}}
  \caption{Generation of Morton-hybrid order}
  \label{fig:morton_hybrid_refinement}
\end{figure}

\begin{figure}[h]
  \centering
     {\includegraphics[width=2.5in]{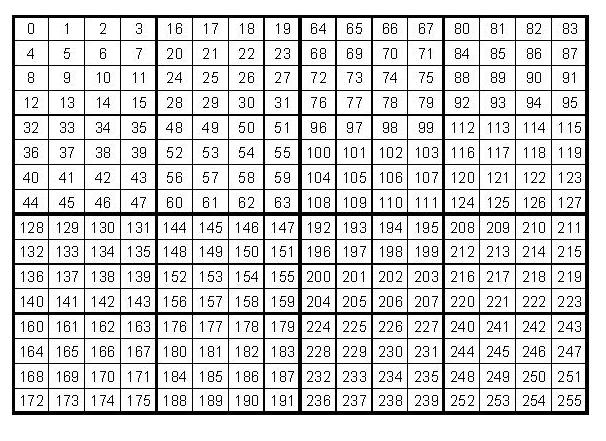}}
  \caption{Matrix in Morton-hybrid Order}
  \label{fig:morton_hybrid_mat}
\end{figure}

To generate the Morton-hybrid order, the procedure for generation of
the Morton order is followed, using the $Z$ pattern as shown in
Fig. \ref{fig:morton_patterns}, until the size of the sub-matrix being
mapped onto memory is $T \times T$. To map the entries of this block
in the one-dimensional array representing the matrix, a row-major
order is assumed. Fig. \ref{fig:morton_hybrid_refinement} shows the
steps in the generation of the $Z$-shaped Morton-hybrid order of an
$16 \times 16$ matrix for truncation size $T=4$. The resulting map
of the entries of the matrix in the corresponding one-dimensional array
in memory can be seen in Fig. \ref{fig:morton_hybrid_mat}.

The following presentation of the Morton-hybrid index for $T=16$ was taken from \cite{Wise:MaskedIntegers}, but we elaborate on it additionally in our Proposition \ref{lowhigh} below and the associated proof. For the remainder of this section, $\Theta$ denotes the Morton-hybrid layout for $T = 16$. We assume the input matrix $M$ has dimension $2^{m} \times 2^{m}$. In Sec. \ref{sec:morton_conv}, we saw that the Morton index is determined using the inter-leaving of the bits of $i$ and $j$, for a given Cartesian index $(i,j)$. In Sec. \ref{sec:row_major_conv} the row major index was composed of the bits of $j$ concatenated to the bits of $i$. The Morton-hybrid order is a combination of both these orders.

Consider the Morton-hybrid matrix shown in Fig. \ref{fig:morton_hybrid_indexing1} showing row-major
sub-blocks of a Morton-hybrid matrix for $T=16$. Each $16 \times 16$
row-major block has a row index and a column index as given in
Sec. \ref{sec:terms}. For example, consider the row-major sub-block
outlined in red in Fig. \ref{fig:morton_hybrid_indexing1}. Its row
index is 1 and its column index is 2. The Morton index of a
sub-block in the Morton-hybrid order is then given by the
interleaving of the block row and block column indices as per the
Morton indexing scheme. Now, each entry $e$ in a Morton-hybrid
matrix can be indexed with the help of information regarding:
\begin{enumerate}
\item{the $T \times T$ row-major sub-block in which it lies}
\item{its row index within this row-major sub-block}
\item{its column index within this row-major sub-block}
\end{enumerate}

For each element $e$ in a $T \times T$ sub-block, the possible
values for the row and column indices of $e$ within this block can
be 0, 1, ..., $T-1$. Thus, the offset values can be represented
using $\beta$ bits where $T = 2^{\beta}$. Let $(i,j)$ denote the
Cartesian index of $e$. We now have:

\begin{prop}\label{prps:index_bits}
The $\beta$ lower order bits of $i$ represent the row index of the
element within the $T \times T$ row-major sub-block and the
$(m-\beta)$ higher order bits of $i$ represent the row index of this
$T \times T$ sub-block.
\label{lowhigh}
\end{prop}
\paragraph{proof}
Let $M$ be a $2^{m} \times 2^{m}$ matrix laid out in the
Morton-hybrid order with truncation size $T=2^{\beta}$. Let $e$ be
any entry in the $T \times T$ row-major sub-block $S_{M}$ of $M$,
and let $(i,j)$ denote the Cartesian index of $e$. First we show
that the bit representations of the row index of the sub-block
$S_{M}$ and the row index of the element $e$ within $S_{M}$ require
$(m-\beta)$ and $\beta$ bits respectively.

The matrix $M$ has dimensions $2^{m} \times 2^{m}$ and the
sub-blocks at the base case have dimensions $T \times T$, with
$T=2^{\beta}$. Thus there are exactly $2^{(m-\beta)}$ rows of $T
\times T$ blocks and the row index $i_{M}$ of a sub-block can be 0,
1,..., $2^{(m-\beta)}-1$. So these row indices require $(m-\beta)$
bits to be represented. The $T \times T$ sub-block $S_{M}$ of $M$ is
in the row-major layout and contains $T$ rows of entries. Let $r$
denote the row of $S_{M}$ in which entry $e$ lies. The row index
$i_{r}$ of $e$ within $S_{M}$, which is the same as the row index of
any entry within $r$, can be $0$, $1$,$\ldots$, $T-1$ and thus its bit
representation requires $\beta$ bits.

Now we identify where to get these bits from. The row index $i$ of
the entry $e$ within $M$ is given by
\begin{equation*}
i = i_{M} \times T + i_{r}
\end{equation*}
which is equivalent to
\begin{equation*}
i = ( i_{M} << \beta ) | i_{r}.
\end{equation*}
As such, $i$ can be represented by $m$ bits: the higher order
$(m-\beta)$ bits are the bits of $i_{M}$, making up the row index of
$S_{M}$ within $M$, and the $\beta$ lower order bits are the bits of
$i_{r}$, making up the row index of $e$ within $S_{M}$. This
concludes the proof.

To illustrate the proposition, take for example the element $(20,6)$ from Fig. \ref{fig:morton_hybrid_indexing1}. In this example, $n=2^{m} =64$, i.e. $m=6$, and $T=2^{\beta}=16$, so $\beta=4$. Consider $i=20 = (010100)_{2}$. The $\beta=4$ lower order bits are $(0100)_{2} = 4$, the element's row index within the row-major block, $i_{r}$. The $m-\beta=2$ higher order bits $(01)_{2} = 2$ identify the row index of the row-major sub-block in which this element lies. The same applies to $j=6=(000110)_{2}$. The $\beta=4$ lower order bits $(0110)_{2}=6$ are the column index of the element within the row-major block. The $m-\beta=2$ higher order bits, $(00)_{2}$, correspond to the column index of the block. 

Now consider a matrix $M$ laid out in the Morton-hybrid order and an entry $e$ in $M$. Let $z_{\Theta}$ denote the Morton-hybrid index of $e$ and $(i,j)$ denote its Cartesian index. Let $S_{M}$ denote the $T \times T$ row-major sub-block in which $e$ lies. Recall the interleaving of the two coordinates $i$ and $j$ presented in Sec. \ref{sec:morton_conv} leading up to the Morton index of $e$. The $T \times T$ sub-blocks of $M$ are laid out in the Morton order, so we use this procedure to inter-leave the $(m-\beta)$ higher order bits of $i$ and $j$ to get the Morton index $z_{M}$ of $S_{M}$. Then, we find the row-major index of $e$ within $S_{M}$: as $S_{M}$
is stored in row-major order, the index within $S_{M}$ of $e$ is in
the context of a row-major ordering scheme. To get the row-major
index of $e$ within $S_{M}$, denoted by $z_{r}$, we concatenate
$j_{r}$ to $i_{r}$, where $i_{r}$ and $j_{r}$ are the row and column
indices of element $e$ within $S_{M}$ respectively (according to
Sec. \ref{sec:row_major_conv}). Because the smallest sub-blocks of $M$
are of size $T \times T$, the Morton-hybrid index of the element $e$
of Cartesian index $(i,j)$ is then given by
\begin{equation*}
( z_{M} \times T^{2} ) + z_{r} \end{equation*} or
\begin{equation*}
( z_{M}<< 2\cdot\beta ) | z_{r}. \end{equation*} Hence, it can be
formed by concatenating the row-major offset of $e$ within $S_{M}$
to the Morton index of $S_{M}$ within $M$. Recall, from
Prop. \ref{prps:index_bits}, that the $(m-\beta)$ higher order bits of
$i$ and $j$ are row and column indices of $S_{M}$ within $M$
respectively and the $\beta$ lower order bits of $i$ and $j$ are the
row and column indices of $e$ within $S_{M}$. That is, if we write
\begin{equation*}
i =
(i_{m-1}i_{m-2}...i_{\beta}i_{\beta-1}i_{\beta-2}...i_{1}i_{0})_{2}
\end{equation*} and
\begin{equation*}
j =
(j_{m-1}j_{m-2}...j_{\beta}j_{\beta-1}j_{\beta-2}...j_{1}j_{0})_{2},
\end{equation*}
the Morton-hybrid index $z_{\Theta}$ is given by
\begin{equation*}\label{eqn:interleaving_mh}
z_{\Theta} =
(i_{m-1}j_{m-1}i_{m-2}j_{m-2}...i_{\beta+1}j{\beta+1}i_{\beta}j_{\beta}i_{\beta-1}i_{\beta-2}...i_{1}i_{0}j_{\beta-1}j_{\beta-2}...j_{1}j_{0})_{2}.
\end{equation*}

\begin{figure}[h]
  \centering
     {\includegraphics[width=6in]{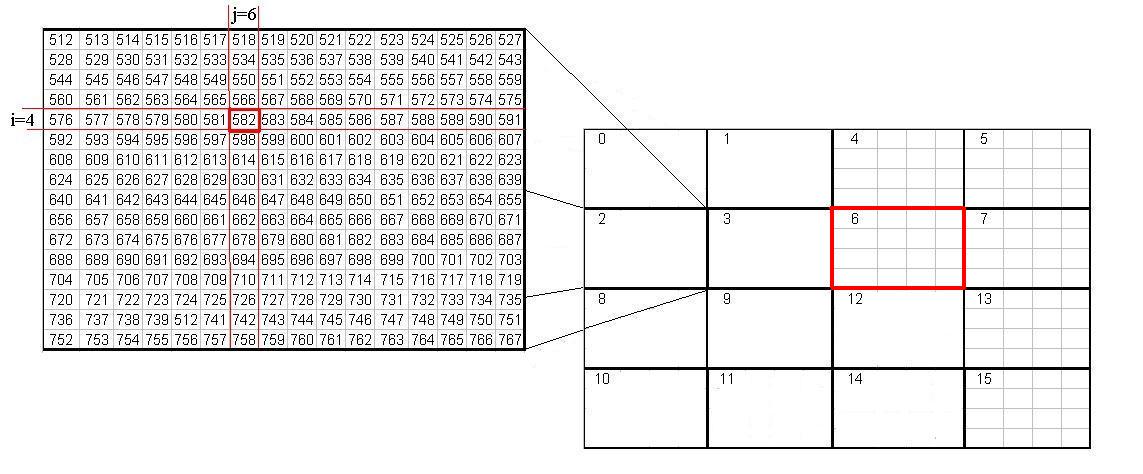}}
  \caption{Morton-Hybrid Ordered Matrix with $T=16$}
  \label{fig:morton_hybrid_indexing}
\end{figure}
\subsubsection{Encoding in the Morton-Hybrid Order}
\label{sec:mh_encoding} To encode a Cartesian index $(i,j)$ in the
Morton-hybrid order, we proceed as follows. The last $\beta$ bits of
$i$ must be extracted. To do this, mask $i$ with \begin{equation*}
\mu = ( 1 << \beta ) - 1
\end{equation*} to extract $i_{r}$,
the part of the binary representation of $i$ representing the row
offset of the element within the row-major block. Thus
\begin{equation*}i_{r} = i \& \mu = (i_{\beta-1}i_{\beta-2}...i_{1}i_{0})_{2}.
\end{equation*}
The rest of $i$ is denoted by $i_{m}$. It represents the row index
of the row-major block in which the element $(i,j)$ lies. The value
of $i_{m}$ is given by
\begin{equation*} i_{m} = i \& ( 0xFFFFFFFF << \beta ) = (i_{m-1}...i_{\beta+1}i_{\beta}0000..)_{2}.
\end{equation*}
We find
\begin{equation*}
j_{r} = (j_{\beta-1}j_{\beta-2}...j_{1}j_{0})_{2}
\end{equation*} and
\begin{equation*}
j_{m} = (j_{m-1}...j_{\beta+1}j_{\beta}0000..)_{2},
\end{equation*} the row-major and Morton parts of the index
$j$ respectively in a similar manner. The values $j_{r}$ and $j_{m}$
are the column offset of the element in the row-major block and the
column index of this block respectively. The Morton-hybrid index
$z_{\Theta}$ is found by dilating $i_{m}$ and $j_{m}$ into
\begin{equation*}
i^{\prime}_{m} = (0i_{m-1}...0i_{\beta+1}0i_{\beta}00000000)_{2}
\end{equation*}
and
\begin{equation*}
j^{\prime}_{m} = (0j_{m-1}...0j_{\beta+1}0j_{\beta}00000000)_{2}
\end{equation*}
respectively using Alg. \ref{alg:dilate_morton}. Then, $z_{\Theta}$ is
given by:
\begin{equation}
\label{eqn:encode_mh} z_{\Theta} = ( ( i^{\prime}_{m} << 1 ) |
j^{\prime}_{m} ) | ( i_{r} << \beta ) | j_{r}.
\end{equation} Note that this equation is in fact made of two parts:
\begin{equation*}
( i^{\prime}_{m} << 1 ) | j^{\prime}_{m},
\end{equation*} which performs the interleaving of the Morton parts of the indices as from
Eq. (\ref{eqn:calc_morton}) from Sec. \ref{sec:morton_conv} and
\begin{equation*}( i_{r} << \beta ) | j_{r}\end{equation*} which
appends the row-major parts of the indices as from
Alg. \ref{alg:encode-row-bit} from Sec. \ref{sec:row_major_conv} with
$m=\beta$.

To verify that this results in $z_{\Theta}$ given in
Eq. (\ref{eqn:interleaving_mh}), we perform:
\begin{equation*}( i^{\prime}_{m} << 1 ) = (i_{m-1}0...i_{\beta+1}0i_{\beta}00...00)_{2},\end{equation*}
\begin{equation*} j^{\prime}_{m} = (0j_{m-1}...0j_{\beta+1}0j_{\beta}00...00)_{2},\end{equation*}
\begin{equation*}( i_{r} << \beta ) =
(i_{\beta-1}i_{\beta-2}...i_{1}i_{0}0...0)_{2},\end{equation*} and
\begin{equation*} j_{r} = (j_{\beta-1}j_{\beta-2}...j_{1}j_{0})_{2}.\end{equation*}
We perform OR operations on the above values to get
\begin{equation*}
z_{\Theta} =
(i_{m-1}j_{m-1}...i_{\beta+1}j_{\beta+1}i_{\beta}j_{\beta}i_{\beta-1}i_{\beta-2}...i_{1}i_{0}j_{\beta-1}j_{\beta-2}...j_{1}j_{0})_{2}
\end{equation*}
as in Eq. (\ref{eqn:interleaving_mh}).

We illustrate this encoding procedure with an example in Appendix \ref{A5}.

\subsubsection{Decoding in the Morton-Hybrid Order}
When decoding an index in the Morton-hybrid order, we start with
\begin{equation*}
z_{\Theta} =
(i_{m-1}j_{m-1}...i_{\beta+1}j_{\beta+1}i_{\beta}j_{\beta}i_{\beta-1}i_{\beta-2}...i_{1}i_{0}j_{\beta-1}j_{\beta-2}...j_{1}j_{0})_{2}
\end{equation*} We wish to find the Cartesian index $(i,j)$ by isolating the bits of
$i$ and the bits of $j$. Recall that $z_{\Theta}$ is in fact made of
three parts as described in the introduction of this section:
\begin{enumerate}
\item{the Morton index of the row-major block in which this element lies}
\item{the row offset of the element within this block}
\item{the column offset of the element within this block}
\end{enumerate} Thus to determine $i$, the Morton index of the
row-major block must be decoded to get a row index, and the row
offset of the element within the block is concatenated to this to
get $i$. The same applies for $j$.

We extract the two parts for $i$ and $j$ from above as follows, in a
reverse process to encoding. Recall from the encoding procedure the
intermediate values $i_{r}$, $i_{m}$, $i^{\prime}_{m}$, $j_{r}$, and
$j_{m}$. We mask $z_{\Theta}$ by
\begin{equation*}( (0xAAAAAAAA << \beta ) << \beta )
\end{equation*} to get $i^{\prime}_{m}$ and
mask $z_{\Theta}$ by \begin{equation*}( (0x55555555 << \beta) <<
\beta )
\end{equation*} to get
$j^{\prime}_{m}$. The result is
\begin{equation*}
i^{\prime}_{m} = (i_{m-1}0...i_{\beta+1}0i_{\beta}000....00)_{2}
\end{equation*}
and
\begin{equation*}
j^{\prime}_{m} = (0j_{m-1}...0j_{\beta+1}0j_{\beta}00...00)_{2}.
\end{equation*}
Then, $(i^{\prime}_{m} >> 1)$ and $j^{\prime}_{m}$ are un-dilated
using Alg. \ref{alg:undilate_morton} to get
\begin{equation*}
i_{m} = (i_{m-1}...i_{\beta+1}i_{\beta}0...0)_{2}
\end{equation*}
and
\begin{equation*}
j_{m} = (j_{m-1}...j_{\beta+1}j_{\beta}0...0)_{2}
\end{equation*} respectively. To extract $i_{r}$ and $j_{r}$,
we mask $z_{\Theta}$ with $( \mu << \beta )$ and $\mu$ respectively
to get \begin{equation*} i_{r} =
(i_{\beta-1}i_{\beta-2}...i_{1}i_{0}0...0)_{2}\end{equation*} and
\begin{equation*}
j_{r} = (j_{\beta-1}j_{\beta-2}...j_{1}j_{0})_{2}.
\end{equation*}
The values for $i$ and $j$ can then be found as:
\begin{equation}
\label{eqn:extract_mh_i}i = i_{m} | (i_{r} >> \beta )
\end{equation}and
\begin{equation}\label{eqn:extract_mh_j}j = j_{m} | j_{r}\end{equation}
resulting in \begin{equation*}
i=(i_{m-1}...i_{\beta+1}i_{\beta}i_{\beta-1}...i_{1}i_{0})_{2}
\end{equation*}
and \begin{equation*}
i=(j_{m-1}...j_{\beta+1}j_{\beta}j_{\beta-1}...j_{1}j_{0})_{2}.
\end{equation*}

We illustrate this decoding procedure with an example in Appendix \ref{A6}.

\subsubsection{Computation Overhead}
As above, assume the input matrix has dimensions $2^{\alpha} \times 2^{\alpha}$, where $\alpha$ is the machine word-size. The encoding algorithm for the Morton-hybrid layout performs the same twenty six bitwise operations as those for the encoding algorithm for the Morton layout for bit dilation, with an additional ten operations: four AND operations, four bit shift operations, and two OR operations. Thus encoding in the Morton-hybrid order costs thirty six bit operations. The decoding algorithm for the Morton-hybrid layout performs the same twenty seven operations used in the decoding procedure for the Morton order. In addition to those, seven bit shift operations, two AND operations, and two OR operations are needed, totalling to eleven additional bit operations. Thus decoding Morton-hybrid index to get the Cartesian index requires thirty eight bit operations.

\section{Summary of findings and a brief note on empirical performance}
\label{PerformAnalysis} 

Our findings so far can be summarised as follows. The overhead for using the Peano layout will be compelling as index conversion invokes operations modulo 3. Whilst the Hilbert layout has been promising for improving memory performance of matrix algorithms in general, and despite that the operations for encoding and decoding in this layout can be performed using bit shifts and bit masks, we will still require $m$ iterations for a $2^{m} \times 2^{m}$ matrix for each single invocation of encoding or decoding. In contrast, we find that the conversions for the Morton and the Morton-hybrid layouts incur a constant number of operations assuming the matrix is of dimensions at most $2^{\alpha} \times 2^{\alpha}$, where $\alpha$ is the machine word-size. For the typical value $\alpha = 64$, such matrix sizes are sufficiently large for many applications. 

The present manuscript is an indispensable precursor for our work in \cite{AA16b}, where we introduce the concepts of {\it alignment} of sub-matrices with respect to the cache lines and their {\it containment} within proper blocks under the Morton-hybrid layout, and describe the problems associated with the recursive subdivisions of TURBO under this scheme. Although the full details of the resulting algorithm are beyond the scope of this paper, we report briefly on experiments that demonstrate how the TURBO algorithm in Morton-hybrid layout attains orders of magnitude improvement in run-time performance as the input matrices increase in size. A more detailed cache analysis is reported in \cite{AA16b}.

We run the serial TURBO algorithm on given matrices stored in the row-major layout and use the index conversion techniques from Sec. \ref{sec:row_major_conv} to complete the row and column permutations. We then run the algorithm on the same matrices stored in the Morton-hybrid order and use the conversion techniques from Sec. \ref{sec:morton_hybrid_conv}. We perform the experiments for a number of different values of $T$, the truncation size at which the Morton Ordering stops and a row-major layout begins. We run our experiments is a Pentium III with processor speed of 800 MHz. Its has 16 KB of L1 cache and 256 KB of L2 cache. It runs a linux operating system of version 2.6.12 with \texttt{gcc} compiler version 4.0.0.

\subsection{Test Cases for Recursive TU Decomposition}

To neutralise the effect of modular aritmetic over finite fields and to be able to exclusively account for the gains induced by the Morton-hybrid order, we generate random $n \times n$ matrices over the binary field. Direct linear algebra over finite fields is an important kernel for several integer factorisation and polynomial factorisation algorithms. We chose to test the Morton-hybrid TURBO algorithm on matrices generated from the Niderretier algorithm for factoring polynomials. These matrices arise from the problem of factoring a polynomial $f$ over the binary field and are given by the equation $N_{f} - I$ where $N_{f}$ is the Niederreiter matrix corresponding to the polynomial $f$ \cite{Niederreiter:A,Niederreiter:C,Niederreiter:B} and $I$ is the identity matrix. If $f$ is a polynomial of degree $n$, then the matrix $N_{f}$ is an $n \times n$ matrix. We vary the value of $n$ from $256$ to $8192$ in the tests. We also vary the truncation size $T$ of the Morton-hybrid layout from $T = $ $16$, $32$, $64$, $128$, $256$, in line with previous Morton-hybrid algorithms for matrix multiplication and Cholesky Factorisation in \cite{Wise:7At1}.

\subsection{Results and Analysis for Recursive TU Decomposition}
\label{sec:results} In this section, we present the performance results for the various truncation sizes we experimented with. Below we summarise the actual run-times for both versions and given truncation sizes. We interpret our results as follows. For small values of $N$, the row-major TURBO beats the Morton-hybrid one. Obviously, the overhead associated with the index conversions required by the Morton-hybrid version during each recursive step dominate the overall run-time for small values of $N$. For all possible truncation sizes, the cross-over point is for $N = 1024$. For larger values of $N$ we actually gain orders of magnitude reduction in overall run-time.  For example, when $N = 2^{13}$, the row major TURBO algorithm concludes within about 38.6 hours, whilst the Morton-hybrid algorithm with truncation size equal to $64$ concludes within $10.6$ hours. Now, for each given value of $N$, the best truncation size seems to be around $T = 32$ and $T = 64$. This is where roughly half of the recursive calls down to a trivial base case block have been dispensed with. For smaller truncation sizes, the loss in performance is due to the recursion overhead. For higher truncation sizes, the loss in performance is associated with poor cache performance. We finally note that not only is the best performance of the Morton-hybrid version is for $T = 32$ and $64$. For those ranges, the rate of deceleration in run-time as $N$ increases is the lowest. 

\begin{table}
  \centering
  \caption{Run-time performance for the Morton-hybrid TURBO}
  \label{runtimes2}
\begin{tabular}{|c|c|}
\hline \textbf{N} & \textbf{Row Major} \\
\hline
\hline
128 & 0.15 sec \\
256 & 1.34 sec  \\
512 & 12.2 sec  \\
1024 & 3.4 min \\
2048 & 28.6 min  \\
4096 & 4.2 hrs  \\
8092 &  38.6 hrs \\
\hline
\end{tabular}
\end{table}

\begin{table}
  \centering
  \caption{Run-time performance for the Morton-hybrid TURBO}
  \label{runtimes}
\begin{tabular}{|c|c|c|c|}
\hline \textbf{T} & \textbf{N} &  \textbf{Morton Hybrid} \\
\hline
\hline
16 & 128 &  0.5 sec \\
16 & 256 &  3.74 sec \\
16 & 512 & 25 sec \\
16  & 1024 & 3 min 12 sec \\
16 & 2048 & 20 min \\
16 & 4096 & 3 hrs\\
16 & 8192  & 21 hrs 24 min \\
\hline
\hline
32 & 128  & 0.4 sec \\
32 & 256  & 2.6 sec \\
32 & 512 & 17 sec \\
32  & 1024  & 2 min \\
32 & 2048  & 12 min 24 sec\\
32 & 4096  & 1 hrs 36 min \\
16 & 8192  & 10 hrs 56 min \\
\hline
\hline
64 & 128  & 0.28 sec \\
64 & 256  & 2 sec \\
64 & 512  & 16 sec \\
64  & 1024  & 1 min 42 sec \\
64 & 2048  & 13 min 18 sec \\
64 & 4096  & 1 hrs 30 min \\
16 & 8192  & 10 hrs 36 min \\
\hline
\hline
128 & 256 & 2 sec \\
128 & 512  & 17 sec \\
128 & 1024  & 2 min 6 sec \\
128  & 2048 & 14 min 12 sec \\
128 & 4096  & 1 hrs 50 min \\
128 & 8192  &  14 hrs \\
\hline
\hline
256 & 512  & 11.67 sec \\
256 & 1024  & 2 min \\
256  & 2048  & 16.3 min \\
256 & 4096  & 1 hr 53 min \\
16 & 8192  &  15 hrs 31 min \\
\hline
\end{tabular}
\end{table}

\subsection{Conclusion}
\label{sec:ind_conv_comparison}

In this paper we have reviewed four major space-filling curve representations as they apply to parallel TU decomposition over finite fields (TURBO). Whilst these representations have been traditionally employed to develop cache-oblivious matrix multiplication and factorisation algorithms, both serial and parallel, we find that they incur additional costs associated with encoding and decoding from the row major layout as needed for the row and column permutations within TURBO. Our detailed analysis of the bit operations required, and in some cases the number of table look-ups, shows that the Morton and Morton-hybrid order are the best candidates. Additionally, the Morton-hybrid order balances this cost with that of recursion overhead and thus is a better candidate than the purely Morton order. The present paper is an indispensable precursor for our work in \cite{AA16b}, where we introduce the concepts of {\it alignment} of sub-matrices with respect to the cache lines and their {\it containment} within proper blocks under the Morton-hybrid layout, and describe the problems associated with the recursive subdivisions of TURBO under this scheme. We develop the full details of a cache oblivious variant of TURBO that observes the alignment and containment of sub-matrices invariably across the recursive steps. The resulting algorithm is inherently nested-parallel, and has low span, for which the natural sequential evaluation order has lower cache miss rate. Our experiments show that the TURBO algorithm in the Morton-hybrid layout attains orders of magnitude improvement in performance as the input matrices increase in size. 


\appendix

\section{Encoding in the Hilbert Order}
\label{A1}

From the example in Fig. \ref{fig:hilbert_indexing} we illustrate how to calculate the Hilbert index $z_{\Theta}$ corresponding to the Cartesian index $(4,6)$. Here, $n=8$ and $m=3$. Write $i=4=(100)_{2} =(i_{2}i_{1}i_{0})_{2}$ and $j=6=(110)_{2}=(j_{2}j_{1}j_{0})_{2}$.

\noindent For iteration $k=0$, we have:
\begin{itemize}
\item{$\rho_{0} = U$ and $z_{0} = 0$}
\item{$v_{0} =(i_{2}j_{2})_{2}=(11)_{2} = 3$}
\item{$\rho_{1}=\mathcal{T_{P}}(\rho_{0},v_{0})= U$}
\item{$z_{1}= (z_{0} << 2) | \mathcal{T_{V}}(\rho_{0},v_{0}) = (10)_{2}$}
\end{itemize}

\noindent For iteration $k=1$
\begin{itemize}
\item{$\rho_{1} = U$ and $z_{1} = (10)_{2}$}
\item{$v_{1} = (\mathcal{T_{P}}(\rho_{1},v_{1})= C$}
\item{$z_{2}= (z_{1} << 2) | \mathcal{T_{V}}(\rho_{1},v_{1}) = (1011)_{2}$}
\end{itemize}

\noindent For iteration $k=2$
\begin{itemize}
\item{$\rho_{2} = C$ and $z_{2} = (1011)_{2}$}
\item{$v_{2} = (i_{0}j_{0})_{2}=(00)_{2} = 0$}
\item{$\rho_{3}=\mathcal{T_{P}}(\rho_{2},v_{2})= C$}
\item{$z_{3}= (z_{2} << 2) | \mathcal{T_{V}}(\rho_{2},v_{2}) = (101110)_{2}$}
\end{itemize}

\noindent We end with $z_{\Theta} = z_{3} = (101110)_{2} = 46$

\section{Decoding in the Hilbert Order}
\label{A2}

We illustrate the process to decode $z_{\Theta} = 46$ from
Fig. \ref{fig:hilbert_indexing}. In this example, $n=8$ and $m=3$.
Start by writing $z_{\Theta} = (101110)_{2}$.

\noindent For iteration $k=0$, we have:
\begin{itemize}
\item{$\rho_{0} = U$, $i_{0} =0$ and $j_{0} = 0$}
\item{$v_{0} =(z_{5}z_{4})_{2}=(10)_{2} = 2$}
\item{$\rho_{1}=\mathcal{T^{\prime}_{P}}(\rho_{0},v_{0})= U$}
\item{$v^{\prime}_{0}=\mathcal{T^{\prime}_{V}}(\rho_{0},{v_{0}})=(11)_{2}$}
\item{$i_{1}= (i_{0} << 1) | (v^{\prime}_{0} >> 1) = (1)_{2}$}
\item{$j_{1}= (j_{0} << 1) | (v^{\prime}_{0} \& 1) = (1)_{2}$}
\end{itemize}

\noindent For iteration $k=1$
\begin{itemize}
\item{$\rho_{1} = U$, $i_{1} = (1)_{2}$, and $j_{1} = (1)_{2}$}
\item{$v_{1} =(z_{3}z_{2})_{2}=(11)_{2} = 3$}
\item{$\rho_{2}=\mathcal{T^{\prime}_{P}}(\rho_{1},v_{1})= C$}
\item{$v^{\prime}_{1}=\mathcal{T^{\prime}_{V}}(\rho_{1},{v_{1}})=(01)_{2}$}
\item{$i_{2}= (i_{1} << 1) | (v^{\prime}_{1} >> 1) = (10)_{2}$}
\item{$j_{2}= (j_{1} << 1) | (v^{\prime}_{1} \& 1) = (11)_{2}$}
\end{itemize}

\noindent For iteration $k=2$
\begin{itemize}
\item{$\rho_{2} = C$, $i_{2} = (10)_{2}$, and $j_{2} = (11)_{2}$}
\item{$v_{2} =(z_{1}z_{0})_{2}=(10)_{2} = 2$}
\item{$\rho_{3}=\mathcal{T^{\prime}_{P}}(\rho_{2},v_{2})= C$}
\item{$v^{\prime}_{2}=\mathcal{T^{\prime}_{V}}(\rho_{2},{v_{2}})=(00)_{2}$}
\item{$i_{3}= (i_{2} << 1) | (v^{\prime}_{2} >> 1) = (100)_{2}$}
\item{$j_{3}= (j_{2} << 1) | (v^{\prime}_{2} \& 1) = (110)_{2}$}
\end{itemize}

\noindent We get $i = i_{3} = (100)_{2} = 4$ and $j=j_{3}=(110)_{2}$
and the resulting Cartesian index is $(4,6)$.

\section{Encoding in the Morton Order}
\label{A3}

We illustrate the process to find the Morton index $z_{\Theta}$
corresponding to the Cartesian index $(4,6)$ from
Fig. \ref{fig:morton_indexing}. We dilate $i = (100)_{2}$ into
$i^{\prime}$ and $j = (110)_{2}$ into $j^{\prime}$ using the
dilation algorithm. This gives $i^{\prime} = (010000)_{2}$ and
$j^{\prime} = (010100)_{2}$. Then $z_{\Theta}$ is found using
Eq. (\ref{eqn:calc_morton}).
\begin{equation*}
z_{\Theta} = (100000)_{2} | (010100)_{2} = (110100)_{2} = 52.
\end{equation*}

\section{Decoding in the Morton Order}
\label{A4}

We illustrate the procedure to find the $(i,j)$ index
corresponding to $z_{\Theta} = 52$ from
Fig. \ref{fig:morton_indexing}. Start by writing $z_{\Theta} =
(110100)_{2}$. This is masked to get
\begin{equation*}
i_{z} = z_{\Theta} \& 0xAAAAAAAA = (100000)_{2}.
\end{equation*}
\begin{equation*}
i^{\prime}_{z} = i_{z} >> 1 = (010000)_{2}
\end{equation*} Then, un-dilating $i^{\prime}_{z}$ gives $i = (100)_{2} = 4$. We
mask $z_{\Theta}$ to get
\begin{equation*}
j_{z} = z_{\Theta} \& 0x55555555 = (010100)_{2}.
\end{equation*}
Un-dilating $j_{z}$ results in $j = (110)_{2} = 6$. Thus the
corresponding Cartesian index is $(4,6)$.

\section{Encoding in the Morton-hybrid Order}
\label{A5}

From the example in Fig. \ref{fig:morton_hybrid_indexing} we illustrate the procedure to calculate the Morton-hybrid index $z_{\Theta}$ corresponding to the Cartesian index $(20,6)$. Here, $T=16$ and $\beta=4$. We write
\begin{equation*}
i = (010100)_{2}
\end{equation*} and
\begin{equation*}
j = (000110)_{2}.
\end{equation*}
Mask these with $\mu = ( 1 << \beta ) - 1$ and $( 0xFFFFFFFF << 4)$
to get $i_{r}$ and $i_{m}$ from $i$ and $j_{r}$ and $j_{m}$ from
$j$. For $\beta=4$, $\mu = (1111)_{2}$. We have:
\begin{equation*}
i_{r} = i \& \mu = (0100)_{2},
\end{equation*}
\begin{equation*}
i_{m} = i \& ( 0xFFFFFFFF << 4 ) = (010000)_{2},
\end{equation*}
\begin{equation*}
j_{r} = j \& \mu = (0110)_{2},
\end{equation*}
and
\begin{equation*}
j_{m} = j \& \mu = (000000)_{2}.
\end{equation*}
Then, $i_{m}$ and $j_{m}$ are dilated into $i^{\prime}_{m}$ and
$j^{\prime}_{m}$ respectively to get:
\begin{equation*}
i^{\prime}_{m} = (000100000000)_{2}
\end{equation*}
and
\begin{equation*}
j^{\prime}_{m} = (000000000000)_{2}
\end{equation*}
Finally, find $z_{\Theta}$ according to Eq. (\ref{eqn:encode_mh}):
\begin{equation*}
z_{\Theta} = ( ( i^{\prime}_{m} << 1 ) | j^{\prime}_{m} ) | ( i_{r}
<< 4 ) | j_{r},
\end{equation*}
resolving to
\begin{equation*}
z_{\Theta} = ( (1000000000)_{2} | (0000000000)_{2} ) |
(01000000)_{2} | (0110)_{2} = (1001000110)_{2}.
\end{equation*} We get $z_{\Theta} = 582$.

\section{Decoding in the Morton-hybrid Order}
\label{A6}

We illustrate this procedure on the matrix in
Fig. \ref{fig:morton_hybrid_indexing}. We start with $z_{\Theta} =
582$ and we aim to extract $i$ and $j$. For this matrix, $T=16$ and
$\beta=4$. Write
\begin{equation*}
z_{\Theta} = (001001000110)_{2}.\end{equation*}We follow the steps
described above. First we extract $i^{\prime}_{m}$ and
$j^{\prime}_{m}$ using the masks $( (0xAAAAAAAA << \beta ) << \beta
)$ and $( (0x55555555 << \beta) << \beta )$ respectively. We get
\begin{equation*}
i^{\prime}_{m} = (001000000000)_{2}
\end{equation*}
and
\begin{equation*}
j^{\prime}_{m} = (00000000000)_{2}.
\end{equation*}
Then un-dilate $(i^{\prime}_{m} >> 1)$ and $j^{\prime}_{m}$ to get
\begin{equation*}
i_{m} = (010000)_{2}
\end{equation*} and
\begin{equation*}
j_{m} = (000000)_{2}
\end{equation*} respectively. We then extract $i_{r}$ and $j_{r}$
from $z_{\Theta}$ as
\begin{equation*}
i_{r} = z_{\Theta} \& (\mu << 4) = (000001000000)_{2}
\end{equation*}
and
\begin{equation*}
j_{r} = z_{\Theta} \& \mu = (000000000110)_{2},
\end{equation*} where $\mu = ( 1 << \beta ) - 1 = (1111)_{2}$, for
$\beta = 4$. We then calculate
\begin{equation*}
i = i_{m} | ( i_{r} >> \beta ) = (010100)_{2}
\end{equation*}
and
\begin{equation*}
j = j_{m} | j_{r} = (000110)_{2}.
\end{equation*}
This gives $i=20$ and $j=6$.

\end{document}